\documentclass[]{jfm}

\usepackage{graphicx}
\usepackage{epstopdf,epsfig}
\usepackage{newtxtext}
\usepackage{newtxmath}
\usepackage{natbib}
\usepackage{hyper ref}
\usepackage{float}
\usepackage{textcomp}
\usepackage{ulem}
\hypersetup{
colorlinks = true,
urlcolor   = blue,
citecolor  = black,
}

\newcommand{\RomanNumeralCaps}[1]
\linenumbers

\title{Experimental study of velocity statistics in wall-bounded turbulent emulsions}

\author{Yaning Fan\aff{1}\footnotemark{},
Yi-Bao Zhang\aff{1}\footnotemark[\value{footnote}]\footnotetext{Author contributions: Y.F. and Y.-B.Z. contributed equally to this work.}, Jinghong Su\aff{1}, Lei Yi\aff{1,2}, Cheng Wang\aff{1,3}
\and Chao Sun\aff{1,4}\corresp{\email{chaosun@tsinghua.edu.cn}}}

\affiliation{\aff{1}New Cornerstone Science Laboratory, Center for Combustion Energy, Key Laboratory for Thermal Science and Power Engineering of Ministry of Education, Department of Energy and Power Engineering, Tsinghua University, 100084 Beijing, China
\aff{2}Department of Physics, University of Massachusetts, Amherst, Massachusetts 01003, USA
\aff{3}ENS de Lyon, CNRS, Laboratoire de physique, F-69342 Lyon, France
\aff{4}Department of Engineering Mechanics, School of Aerospace Engineering, Tsinghua University, Beijing 100084, China}

\begin{document}
\maketitle

\begin{abstract}
Turbulent emulsions are ubiquitous in chemical engineering, food processing, pharmaceuticals, and other fields. However, our experimental understanding of this area remains limited due to the multi-scale nature of turbulent flow and the presence of extensive interfaces, which pose significant challenges to optical measurements. 
In this study, we address these challenges by precisely matching the refractive indices of the continuous and dispersed phases, enabling us to  measure local velocity information at high volume fractions.
The emulsion is generated in a turbulent Taylor-Couette flow, with velocity measured at two radial locations: near the inner cylinder (boundary layer) and in the middle gap (bulk region). Near the inner cylinder, the presence of droplets suppresses the emission of angular velocity plumes, which reduces the mean azimuthal velocity and its root mean squared fluctuation. The former effect leads to a higher angular velocity gradient in the boundary layer, resulting in greater global drag on the system. In the bulk region, although droplets suppress turbulence fluctuations, they enhance the cross-correlation between azimuthal and radial velocities, leaving the angular velocity flux contributed by the turbulent flow nearly unchanged. In both locations, droplets suppress turbulence at scales larger than the average droplet diameter and increase the intermittency of velocity increments. However, the effects of the droplets are more pronounced near the inner cylinder than in the bulk, likely because droplets fragment in the boundary layer but are less prone to breakup in the bulk. Our study provides experimental insights into how dispersed droplets modulate global drag, coherent structures, and the multi-scale characteristics of turbulent flow.

\end{abstract}

\begin{keywords}
	Emulsions, Multiphase Flow, Taylor-Couette Flow
\end{keywords}

\section{Introduction}
\label{sec:Intro}

Emulsions, consisting of two immiscible liquids, are ubiquitous in various natural and industrial processes, such as oil spills in the ocean, pharmaceuticals, food processing, oil production, and recovery \citep{li1998relationship,PhysRevLett.104.054501,spernath2006microemulsions,mcclements2004food,kokal2005crude,mandal2010characterization,kilpatrick2012water}. In turbulent emulsions, the multi-scale nature of turbulence adds to the complexity of the system. Therefore, although emulsions are frequently encountered, our understanding of the underlying physics of emulsions, particularly turbulent emulsions, remains elementary. 


Two major unanswered questions in the study of emulsions pertain to 1) droplet dynamics (the droplet size distribution and preferential concentration of droplets in certain regions of the flow) and 2) the modulation of the statistics of the base flow. The investigation on droplet size in turbulent flow can be traced back to \citet{kolmogorov_disintegration_1949} and \citet{hinze_fundamentals_1955}. In their theory, the background turbulence is assumed to be homogeneous and isotropic, with the maximum droplet size, $d_H$, determined by the competition between turbulent fluctuations induced external pressure force from the background flow over the droplet and resisting capillary cohesive forces of the droplets \citep{risso1998oscillations,perlekar_droplet_2012,eskin2017modeling,perlekar_spinodal_2014,rosti_droplets_2019,begemann2022effect}.
However, it has been pointed out that the KH theory has certain limitations, particularly in non-homogeneous turbulent flows \citep{lemenand2017turbulent,hinze_fundamentals_1955}. 
Recent studies on wall-bounded Taylor-Couette turbulence \citep{yi_physical_2022} suggest that the average droplet size in the bulk region is not governed by the Kolmogorov-Hinze model \citep{kolmogorov_disintegration_1949,hinze_fundamentals_1955}, but instead by the dynamic pressure induced by the gradient of the mean flow, as proposed by Levich \citep{levich1962physicochemical}.
For more details on the breakup of bubbles and droplets in turbulence, we refer to the review paper by \cite{ni2024deformation} and the references therein.

In turn, the presence of droplets can also alter the statistical properties of turbulence and the global transport quantities, such as drag, in the emulsion \citep{piela2008phase,yi_global_2021}.
\citet{yi_global_2021,yi_physical_2022} and \cite{wang_turbulence_2022} found that the size and viscosity of the droplets have a profound effect on the global drag of the system. The drag increases with the volume fraction and viscosity of the dispersed phase, and the effective viscosity of the emulsion system exhibits a shear-thinning effect as the shear increases. \citet{bakhuis_catastrophic_2021} and \citet{yi2024divergence} studied catastrophic phase inversion in turbulent Taylor-Couette flow, demonstrating that significant drag reductions in the transport of emulsions can be achieved by selecting the appropriate emulsion type. 
The modulation of turbulence by the presence of droplets in turbulent emulsions has recently attracted growing interest in numerical simulations. For small-scale statistics, \citet{crialesi-esposito_modulation_2022} analyzed velocity and vorticity fluctuations and observed a significant deviation in the tails of probability density functions. The observation that a deformable interface increases intermittency has been further verified, with the effect primarily attributed to the strong velocity differences across the interfaces \citep{crialesi-esposito_intermittency_2023}. For large-scale modulation, \cite{hori_interfacial-dominated_2023} reported two regimes in Taylor-Couette emulsion at a Reynolds number of 960 based on the Weber number: an advection-dominated regime and an interface-dominated regime, where both the global drag and Taylor roll structures are strongly modulated. More specifically \citet{dodd_interaction_2016} and \citet{perlekar2019kinetic} examined the turbulent kinetic energy (TKE) budget for varying Weber numbers in homogeneous isotropic turbulence. Both studies found that the total kinetic energy compensates for surface area variations; i.e., the presence of interfaces introduces an alternative kinetic energy transfer mechanism, with TKE decreasing as the interface increases, and vice versa. \citet{trefftz-posada_interaction_2023} conducted a similar investigation in homogeneous shear turbulence, analyzing production, dissipation, and surface tension terms separately to study their modulation effects on TKE. They proposed `catching-up' mechanisms to explain the higher degree of enhancement of production and dissipation rate of TKE for lower Weber number case.

\begin{figure}
	\centerline{\includegraphics[width = 0.95\textwidth]{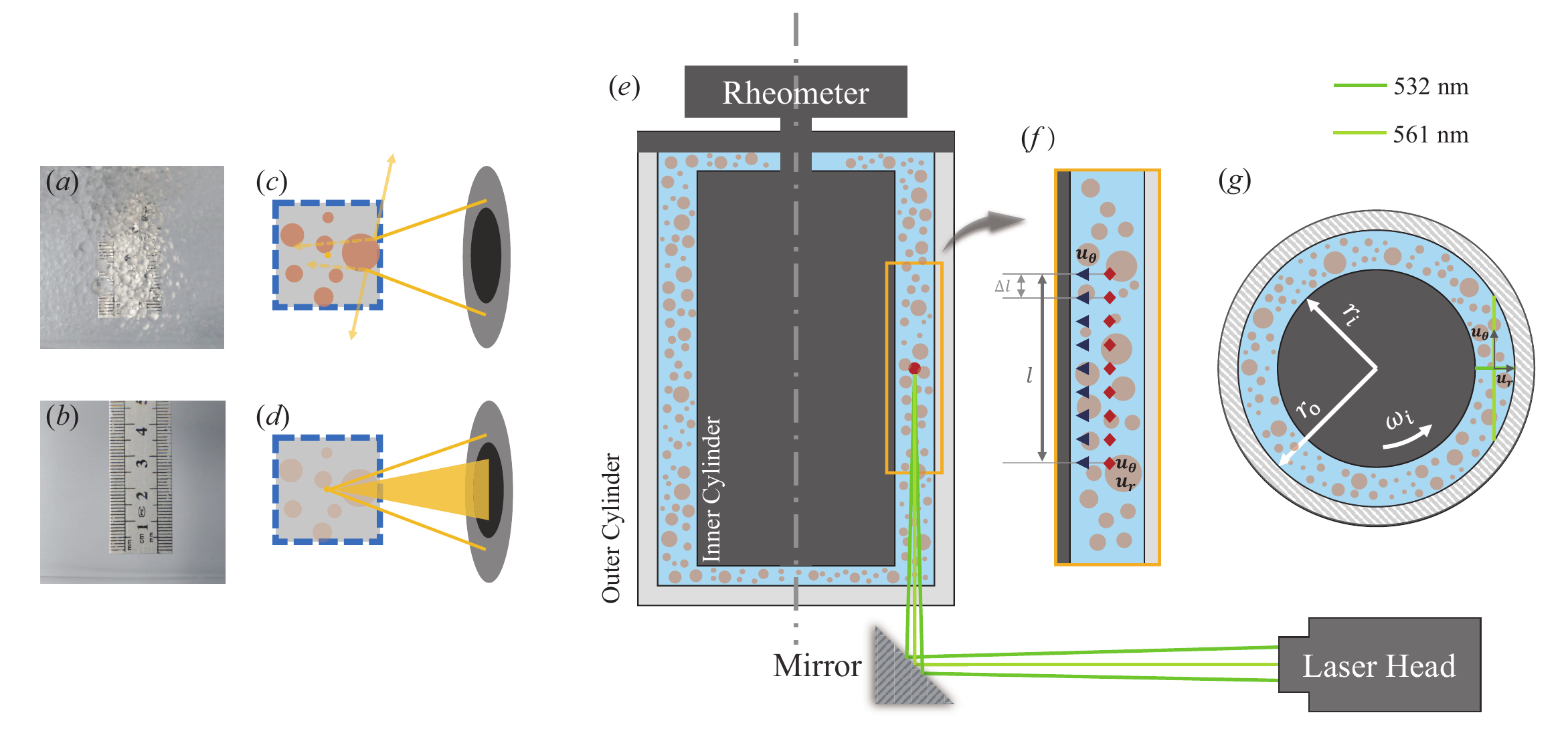}}
	\caption{A picture of a ruler immersed in an emulsion fluid before (\textit{a}) and after (\textit{b}) refractive index matching. The cartoon illustrates the Laser Doppler Anemometer measurements before (\textit{c}) and after (\textit{d}) refractive index matching. We note that in (\textit{d}), the light brown circles are included to remind the reader that the flow system consists of droplets, which are not actually visible. (\textit{e}) A sketch of the experimental setup, with the measurement technique LDA also depicted. (\textit{f}) An enlarged view of the LDA measurement location: the blue triangles represent measurement points near the inner cylinder, while the red diamonds represent points in the middle gap. (\textit{g}) A cross-section of the TC system, with the inner and outer cylinder radii denoted as $r_i$ and $r_o$, respectively. The measured velocity components in the azimuthal and radial directions are denoted as $u_\theta$ and $u_r$, respectively. We note that the circles are shown to remind the reader that liquid-liquid two-phase turbulence is being investigated here; these droplets cannot actually be seen.}
    \label{fig:rim_setup}
\end{figure}

From the above, it is clear that most experimental studies were conducted on the global properties of emulsions when concerning about system modulations, and our knowledge of turbulence modulation by droplets based on experiments is limited. Experimental techniques using optical measurements have been widely applied in fluid dynamics to provide detailed velocity information with high spatial and temporal resolution. However, a major challenge arises for optical measurements in multiphase flows, especially emulsions, because the emulsion fluid is opaque due to the presence of liquid-liquid interfaces. This is illustrated in Fig. \ref{fig:rim_setup}(\textit{a}), where a ruler is immersed in the fluid as a reference. Although the dispersed and continuous phases are individually transparent, their mixture is not. The small droplets in the emulsion system act as spherical lenses. When a light beam passes through the mixture, it experiences multiple reflections and refractions (see Fig. \ref{fig:rim_setup}(\textit{b})), making the ruler difficult to see and rendering optical measurements unfeasible. In light of this challenge, previous optical measurements in liquid-liquid flows were primarily conducted under stratified conditions, at low Reynolds numbers, or at low volume fractions of the dispersed phases \citep{conan2007local,kumara2010particle,morgan2013characteristics,ibarra_dynamics_2018,ibarra_experimental_2021, yi2023recent}. Thus, our experimental understanding of the underlying mechanisms of turbulence modulation in turbulent emulsions remains limited, particularly at high dispersed volume fractions. 

In this work, we address this challenge by precisely matching the refractive indices of the dispersed and continuous phases and experimentally investigating the mechanisms of drag and turbulence modulation by the dispersed droplets. This paper is organized as follows: In \S \ref{sec:Method}, we introduce the experimental setup, the unique preparation of working fluids, and the methodology utilized in detail. In \S \ref{sec:result}, we present the major results, including the global drag modulation and the modification of the continuous phase at two radial positions. Finally, in \S \ref{sec:conclusion}, we conclude the paper with a brief summary and outlook.

\section{Experimental setup and methodology}

\label{sec:Method}
\subsection{Experimental setup}
\label{sec:Exp_setup}

Taylor-Couette (TC) turbulence, which refers to the flow between two coaxial cylinders, is one of the classic systems in fluid mechanics. TC is a closed system that allows for precise control of the volume fraction of the dispersed phase and features easy accessibility for global drag and local velocity measurements. In our experiments, the TC system is constructed from a commercial rheometer (Discovery Hybrid Rheometer, TA Instruments), as shown in Fig. \ref{fig:rim_setup}(\textit{e}). The system consists of an inner cylinder with a radius $r_i = 25$ mm and height $L = 100$ mm, and an outer cylinder with a radius $r_o = 35$ mm and height $110$ mm. These two cylinders create a gap $d = r_o - r_i = 10$ mm, a radius ratio $r_i/r_o = 0.714$, and an aspect ratio $\Gamma = L/d = 10$. The inner cylinder is made of aluminum and anodized to form a black oxidation layer, which reduces unwanted reflection that could decrease the signal-to-noise ratio during optical measurements. The outer cylinder and the cubic tank, which encloses the outer cylinder, are made of plexiglass, allowing for optical accessibility. The upper cover is made of black Acrylonitrile Butadiene Styrene (ABS) to further minimize unwanted reflection from above.


The inner cylinder is directly screwed onto the rheometer and driven by its motor at a constant angular velocity $\omega_i$. The outer cylinder, associated with the cubic tank, is fixed. The global torque required to maintain a constant $\omega_i$ can be directly measured by the rheometer, with high accuracy of up to 0.1 nN$\cdot$m. The time series of torque is recorded after a statistically steady state is reached. The measured torque consists of two parts: one contributed by the TC flow and the other originating from the top and bottom end plates (von K\'{a}rm\'{a}n flow). The latter contribution can be estimated using the same linearization method that has been fully discussed in our previous studies \citep{yi_physical_2022,wang_how_2022}. The torque contributed by the TC flow is denoted as $T$ in the following.

The control parameter of the TC system is the Reynolds number defined as 
\begin{equation}
	\Rey  = \omega_i r_i d/\nu_c. \label{equ:Re}
\end{equation}
where $\nu_c=\mu_c/\rho_c$ is the kinematic viscosity of the continuous phase. $\nu$, $\mu$, and $\rho$ are the kinematic viscosity, dynamics viscosity, and density of the fluid. We use subscripts $c$ and $d$ to discriminate the continuous and dispersed phase. For the two phase case, volume fraction of the dispersed phase $\phi$ is needed. 
A key response parameter of the system is the dimensionless torque defined as
\begin{equation}
	G = T/(2\pi L \rho_c \nu_c^2).  \label{equ:G}
\end{equation}

\subsection{Working fluid}

As mentioned in the introduction, the presence of a two-phase interface typically makes the fluid optically opaque, which greatly limits the applicability of optical measurements. To address this issue, we precisely match the refractive index between the continuous and dispersed phases \citep[see review articles][]{wright_review_2017,amini_investigation_2012,wiederseiner_refractive-index_2011,budwig_refractive_1994}. After refractive index matching (RIM), the droplets are invisible within the continuous phase, making the entire emulsion system optically transparent, as sketched in Fig. \ref{fig:rim_setup}(\textit{c,d}).

In this work, the continuous phase is a mixture of ultra-pure water (density $\rho_w = 0.998 \times 10^3 \ \mathrm{kg/m^3}$, refractive index $n_w = 1.3325$) and glycerol (Titan Greagent, G66258A, $\rho_g = 1.231 \times 10^3 \ \mathrm{kg/m^3}$, $n_g = 1.4720$). The dispersed phase is a mixture of silicone oil (Shin-Etsu KF-96L-2cSt, $\rho_s = 0.873 \times 10^3 \ \mathrm{kg/m^3}$, $n_s = 1.391$) and ethoxy-nonafluorobutane (Novec 7200 3M\texttrademark, $\rho_n = 1.420 \times 10^3 \ \mathrm{kg/m^3}$, $n_n = 1.3014$). Here, $n$ denotes the refractive index of the fluid. To match the refractive index, we first estimate the volume fraction of the $i$-th component, $\varphi_i$, according to the classic Newton's equation
\citep{newton_opticks_1704,kurtz_refractivity_1936,reis_refractive_2010}:
\begin{equation}
	n_{12}^2 = \varphi_1 n_1^2 + \varphi_2 n_2^2, \label{equ:index}
\end{equation}
where $n_i$ is the refractive index for the $i$-th component and $n_{12}$ is for the mixture. We measure the refractive indices of the continuous and dispersed phases, $n_c$ and $n_d$, using a commercial refractometer (GR30, Shanghai Zhuoguang Instrument Technology Co., Ltd.) with a resolution of $0.0001$ and an accuracy of $\pm 0.0002$ at a temperature of 25 \textcelsius, which is the temperature of the emulsion during the experiment. We then compare $n_c$ and $n_d$ and carefully adjust $\varphi_{1,2}$ until $n_c = n_d$ within the precision of the refractometer. It is important to note that even the addition of 0.1 g of silicone oil or Novec 7200 to a 200 g mixture of oil and Novec 7200 will alter its refractive index. Additionally, the densities of the two phases must also be matched to eliminate the effect of centrifugal force \citep{yi_global_2021}. If the densities are not matched, we have to adjust the target refractive index and repeat the procedure described above until both refractive index matching (RIM) and density matching are satisfied. {The volume fractions of each fluid for the continuous phase are $\phi_w \approx 87.9 \%$ and $ \phi_g \approx 12.1 \% $, and those for the dispersed phase are $\phi_s \approx 65.9\% $ and $ \phi_n \approx 34.1\%$. }


Consequently, the water-glycerol (W\&G) mixture, as the continuous phase, has a density of $\rho_c = 1.033 \times 10^3 \ \mathrm{kg/m^3}$, kinematic viscosity of $\nu_c = 1.26 \times 10^{-6} \ \mathrm{m^2/s}$, and refractive index of $n_c = 1.3506$. The silicon-oil-Novec (O\&N) mixture, as the dispersed phase, has $\rho_d = 1.046 \times 10^3 \ \mathrm{kg/m^3}$, $\nu_d = 1.15 \times 10^{-6} \ \mathrm{m^2/s}$, and $n_d = 1.3506$. The interfacial tension of the two phases is calculated based on the equation proposed by \citet{girifalco_theory_1957} for solid-liquid systems, which is equally valid for liquid-liquid systems \citep{lee_scope_1993}
\begin{equation}
	\gamma=\gamma_c+\gamma_d-2\zeta(\gamma_c \gamma_d)^{1/2}, \label{equ:interfacetension}
\end{equation}
where $\gamma_{c}$ ($\gamma_{d}$) is the surface tension of the continuous (dispersed) phase, and $\gamma$ is the interfacial tension between the two phases. $\zeta$ is the interfacial interaction parameter, set to a value of 1. The values of $\gamma_c = 73.28 \times 10^{-3} \ \mathrm{N/m}$ and $\gamma_d = 15.73 \times 10^{-3} \ \mathrm{N/m}$ are measured in air using the pendant drop method. The interfacial tension between W\&G and O\&N is $\gamma = 21.11 \times 10^{-3} \ \mathrm{N/m}$. {In turbulent emulsions, dynamic pressure and viscous stress deform and break up the droplets, while interfacial tension resists deformation. The effect of dynamic pressure is measured by Weber number of the system \citep{crialesi-esposito_interaction_2023,hori_interfacial-dominated_2023,su_numerical_2024,su_turbulence_2024,Su_Zhang_Wang_Yi_Xu_Fan_Wang_Sun_2025}, defined as $We_{sys} = \rho_c u_\tau ^2 d /\gamma$, is $We_{sys} = 4$. Here, $u_\tau = \sqrt{T/(2\pi r_i^2 L \rho_c)}$ is the friction velocity. The effect of viscous stress can be measured by the capillary number  $Ca = \tau_\nu/\tau_\gamma = {\nu \rho_c \epsilon^{1/2} D}/(\gamma \nu^{1/2})$, where the average energy dissipation rate $\epsilon$ is estimated from the global torque $\epsilon \approx 0.1T \omega_i/(\pi (r_o^2-r_i^2)L\rho_c)$ \citep{ezeta2018turbulence,yi_physical_2022}. In our study, the Capillary number is approximately $0.01$, which is much smaller than $1$. Therefore, viscous stress is not important in determining the droplet deformation and breakup.} 



\subsection{Velocity measurement in the continuous phase}

We employ a laser Doppler anemometer (LDA, TSI) to measure velocity due to its high temporal and spatial resolution. The LDA consists of two pairs of laser beams, with wavelengths of 561 nm and 532 nm, as shown in right part of Fig. \ref{fig:rim_setup}. We use the 561 nm and 532 nm laser beams to measure the azimuthal and radial velocity components, denoted as $u_\theta$ and $u_r$, respectively  (see Fig. \ref{fig:rim_setup}(\textit{g})). The beam waist diameter and length of the measurement volume are around $260$ \textmu m and $3.2$ mm, respectively. The physical size of the LDA measurement volume will not affect the measured data and the details are discussed in appendix \ref{sec:append_data_process}. The laser head of the LDA is mounted on a high-precision platform, which can move independently in the \textit{x}, \textit{y}, and \textit{z} directions. It consists of two parts: a vertical electric lifting table (Fly-opt, PSTV50-S57) with a precision of 20 \textmu m and a maximum stroke of 50 mm, and an \textit{x-y} electric horizontal displacement table (RedStarYang EPSB-150-B-G and EPSB-50-B-G) with precisions of 20 \textmu m and 10 \textmu m, and maximum strokes of 150 mm and 50 mm in the \textit{x} and \textit{y} directions, respectively. This platform allows us to change the measurement location. A $45^\circ$ mirror is used to reflect the laser beam to pass through the TC system from below. In this way, distortion of the laser beam due to the curved outer cylinder can be avoided \citep{huisman_applying_2012}. The seeding particles used are polystyrene (Ruige Tech.) with an average diameter of 5 \textmu m and a density of $1.05 \times 10^3 \ \mathrm{kg/m^3}$. The surfaces of the polystyrene particles are grafted with hydroxyl groups to make them hydrophilic. Thus, the particles can only dissolve in the continuous phase, whose velocity information can be measured by the LDA.

In LDA measurements, a slight mismatch in the refractive index or variation in temperature from 25 {\textcelsius} can cause the data rate to drop sharply. Therefore, precisely matching the refractive index and controlling the temperature are crucial. Temperature-controlled water from a refrigerator is circulated between the outer cylinder and the cubic tank. A PT100 thermocouple is used to measure the temperature of the emulsion, with its variation from 25 {\textcelsius} remaining within $0.1$ {\textcelsius} during the experiment. We also finely tune the concentration of the seeding particles to ensure that the data rate of the LDA in both single-phase and two-phase measurements is as close as possible.

{The TC flow can be divided into two parts: the boundary layer near the inner and outer cylinder where there exist high velocity gradient, and the bulk region
where the flow is nearly homogeneous and isotropic \citep{grossmann_highreynolds_2016}.} Therefore, in our experiments the velocity measurements are performed at two representative locations in the radial direction: near the inner cylinder and at the middle gap. In the former case, the dimensionless distance from the inner cylinder is given by $\widetilde{r} = \frac{r-r_i}{r_o-r_i} = 0.05$, and in terms of the viscous length scale, $y^{+}=\frac{r-r_i}{\delta_\nu} \approx 35$ \citep{huisman_logarithmic_2013}. {The viscous length $\delta_\nu$ is defined as $\delta_\nu = \nu_c / \sqrt{T/(2\pi r_i^2 L \rho_c)}$. It is based on the properties of continuous phase considering the minor differences between two phases. The torque $T$ used for $\delta_{\nu}$ estimation is taken from the single-phase case since the global drag is only slightly enhanced.} In the latter case, the dimensionless distance from the inner cylinder is $\widetilde{r} = 0.5$. At the middle gap, both $u_\theta$ and $u_r$ are measured, while only $u_\theta$ can be obtained near the inner cylinder due to the obstruction of the laser beam by the bottom end of the inner cylinder. 
{For each radial position, by adjusting the location of the laser head every 2 mm, LDA measurements are taken at 10 different heights, separated by $\Delta l \approx 2.70$ mm due to the difference in refractive indices of air and the working fluid. The axial length of one pair of Taylor vortices is around $2.1d$ \citep{Zhang_Fan_Su_Xi_Sun_2025}. Thus, velocity data at 9 axial heights, spanning a length of $l \approx 21.63$ mm $\approx 2.1 d$, are used with their center located at the middle height of the TC setup. In this way the influence of secondary flow can be minimized. } 
{For the LDA technique we use in this work, the total amount of data points acquired once can be set in the software. Near the inner cylinder, a fixed amount of data points of $5 \times 10^5$ is acquired since only $u_\theta$ is collected. While at the middle gap, the total amount of data points for two velocity components is set to be $1 \times 10^6$. The actual amount of data points for $u_\theta$ and $u_r$ is determined by the average sampling frequency, i.e. data rate, while the total measurement time is the same. Higher sampling frequency in one direction, more data points in that direction and fewer in the other. However, since the data rates of two components are quite close, the amount of data points in both directions is approximately $5 \times 10^5$.}
In this way, the velocity components $u_\theta$ and $u_r$ of the continuous phase, which are functions of vertical position $z$ and time $t$, can be obtained. The details of pre-processing of velocity time series are introduced in appendix \ref{sec:append_data_process}. Since TC turbulence is statistically steady, the velocity components can be decomposed into the mean and fluctuation parts. Taking $u_\theta$ as an example, $u_\theta = \langle u_\theta \rangle_{t} + u_\theta '$, where $\langle\ \rangle_{t}$ denotes averaging over time. In TC turbulence, due to the presence of Taylor vortices, turbulence statistics depend on the vertical position. We thus also perform an additional average in the $z$ direction, i.e., $\langle u_\theta \rangle_{t,z}$. {The root mean squared (rms) fluctuation of $u_\theta$ is defined as $\sigma(u_\theta ') =\langle \sqrt{\langle u_\theta '^2 \rangle_{t}}\rangle_{z}$.}

\begin{figure}
	\centerline{\includegraphics[width = 0.6\textwidth]{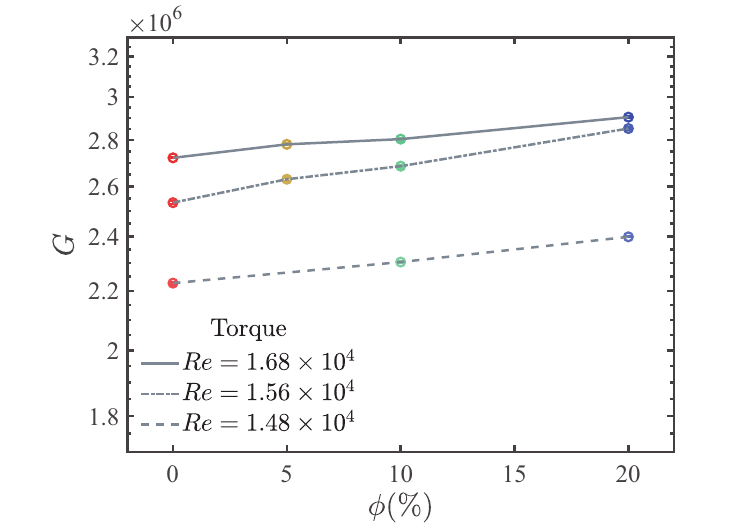}}
	\caption{The dimensionless torque, $G$, is presented as a function of the volume fraction of the dispersed phase for three different Reynolds numbers, $\Rey$. The datasets at $\Rey = 1.68 \times 10^4$ and $\Rey = 1.48 \times 10^4$ are measured in this work, while the dataset at $\Rey = 1.56 \times 10^4$ is from \cite{yi_global_2021}.}
	\label{fig:drag}
\end{figure}

To measure the droplet size, Oil Red O (MCE) is dissolved in the droplets to make them visible.  The illumination of the droplets is achieved through backlighting, and their images are captured by a high-speed camera (Photron NOVA S12) equipped with a micro lens. Considering that the size of the droplets and the measurement window are much smaller than the diameter of the outer cylinder, the distortion of the droplet image due to curvature can be neglected \citep{yi_global_2021}. The average droplet size is computed from approximately $10^3$ droplets. {We conduct the diameter measurement at $\Rey = 1.68 \times 10^4$ with a volume fraction $\phi=1\%$ in the bulk region of TC flow. We use the symbol $\langle D \rangle$ to denote the mean diameter, which is $\langle D \rangle \approx 700$ \textmu m. Further discussions on the mean droplet size and its radial dependence are included in the appendix \ref{sec:append_diameter}.}

\section{Results and discussion}
\label{sec:result}

The dimensionless torque of the inner cylinder defined in Eq. \ref{equ:G} for different O\&N volume fractions, $\phi$, and Reynolds number, \Rey, is depicted in Fig. \ref{fig:drag}. We also compare our results with previous study \citep{yi_global_2021}, shown as dashed-dotted line in Fig. \ref{fig:drag}. It is found that $G$ increases with $\phi$ for a fixed \Rey, consistent with our previous studies (\citet{yi_global_2021,yi_physical_2022,wang_turbulence_2022}). {Note that the slope of $G$ obtained in previous work is different from the current one. This may be caused by the different interfacial tension, which would influence the droplet size and its distribution. According to \cite{su_numerical_2024}, the interface contribution to the drag is a function of interfacial tension and curvature of the droplet interface.} The increase of the global drag signifies that the turbulent flow is modified by the presence of droplets, {not only the global transportation but also the local flow statistics, such as the mean velocity and rms velocity fluctuation}. In the following, we reveal how droplets modulate the drag and the turbulence of the continuous phase. 

\subsection{Modulation of continuous phase near the inner wall}
\label{subsec:IW}

\begin{figure}
	\centerline{\includegraphics[width = 0.95\textwidth]{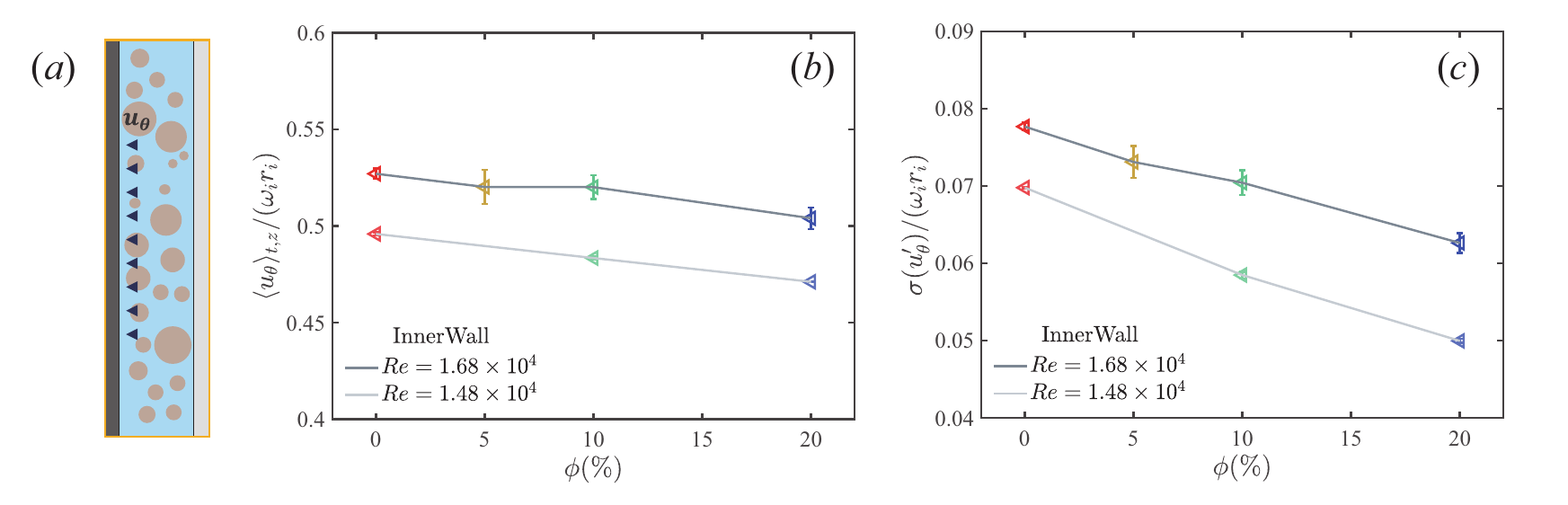}}
	\caption{(\textit{a}) The schematic shows the LDA measurement positions near the inner cylinder, represented by the blue triangles. Only the azimuthal velocity $u_\theta$ can be measured here due to the shielding of the laser beams by the inner cylinder. Average velocity (\textit{b}) and root mean squared velocity fluctuation (\textit{c}) of $u_\theta$ are presented with varying volume fraction $\phi$ for two Reynolds numbers. Both $\langle u_\theta \rangle_{t,z}$ and $\sigma(u_\theta^\prime)$ are normalized by the velocity of the inner cylinder. }	
	\label{fig:IW-fluc}
\end{figure}

We start with the turbulence statistics near the inner cylinder ($\Tilde{r}=0.05$, $y^+=35$), where only the azimuthal velocity is measured as depicted in Fig. \ref{fig:IW-fluc}(\textit{a}). The average velocity $\langle u_\theta \rangle_{t,z}$ as well as rms fluctuation $\sigma(u_\theta^\prime)$ with varying O\&N volume fraction $\phi$ at two different Reynolds numbers $\Rey$ are shown in Fig. \ref{fig:IW-fluc}(\textit{b,c}). {Note that the error bars are calculated based on the following steps:  We first use bootstrap method to calculate the mean value and error bar for a given quantity at one axial height; the mean values and error bars are then averaged over one pair of Taylor vortices.} Apparently, for $u_\theta$ near the inner cylinder, both $\langle u_\theta \rangle_{t,z}$ and $\sigma(u_\theta^\prime)$ decrease with volume fraction. The percentage of decrease for $\sigma(u_\theta^\prime)$ can even reach up to approximately 20\%, indicating a strong suppression of the turbulence intensity near the inner cylinder with the presence of droplets. 

\begin{figure}
	\centering
	\includegraphics[width=0.95\textwidth]{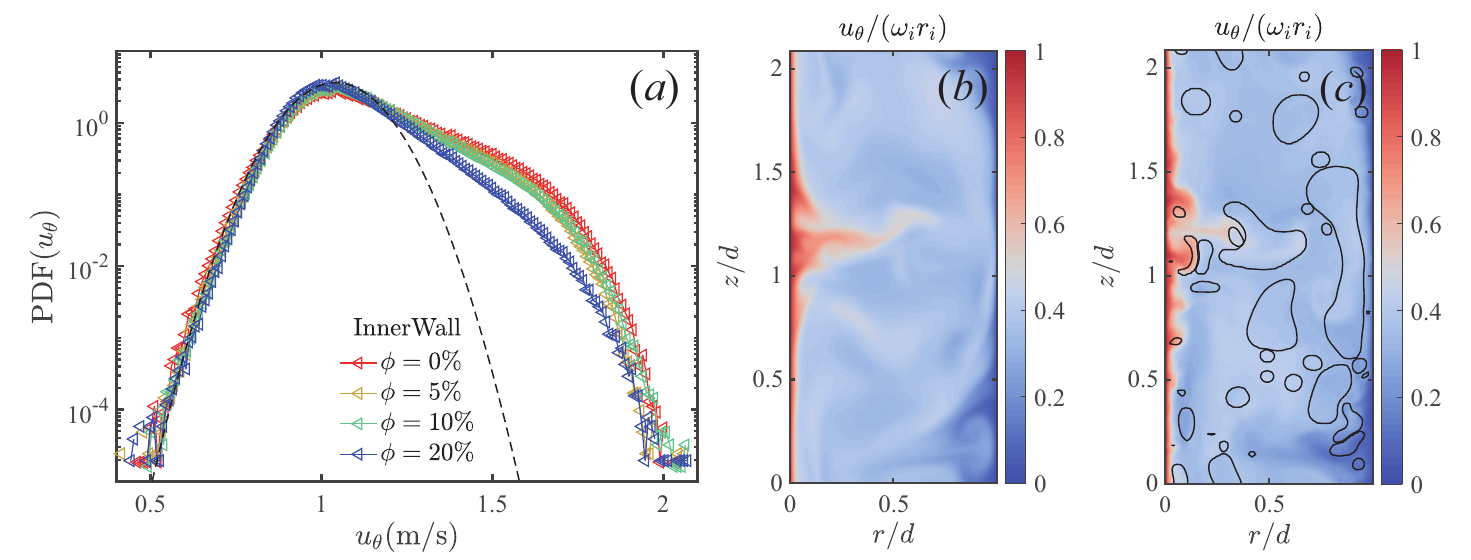}
	\caption{(\textit{a}) Probability density functions of the azimuthal velocity $u_\theta$ near the inner cylinder for varying volume fractions $\phi$ at $Re = 1.68 \times 10^4$. The black dashed line represents the Gaussian distribution fitted to the left part of the data. The contour plots of the instantaneous azimuthal velocity field $u_\theta$ for the single-phase case (\textit{b}) and the two-phase case with $\phi=20\%$ (\textit{c}) at $\Rey = 0.52\times 10^4$. The solid lines in (\textit{c}) indicate the interfaces of the droplets. {Results in (\textit{b}) and (\textit{c}) are from the numerical simulation \citep{su_numerical_2024}.}}
	\label{fig:pdf&simulation}
\end{figure}

To provide further insight, we analyze the probability density functions (PDFs) of $u_\theta$ at $\Rey = 1.68 \times 10^4$ for different volume fractions $\phi$, as depicted in Fig. \ref{fig:pdf&simulation}(\textit{a}). While the left tails can be fitted with a Gaussian distribution (see the dashed line in Fig. \ref{fig:pdf&simulation}(\textit{a})), the right tails are non-Gaussian. 
The overall PDFs are positively skewed. With the presence of droplets, the right tails shrink as the volume fraction increases. In Rayleigh-B\'{e}nard (RB) turbulence, a nearly identical shape has been reported for the temperature PDF \citep{emran_fine-scale_2008}. This PDF is induced by a combination of bursting plumes detaching from the thermal boundary layer and large-scale rolls \citep{castaing_scaling_1989,yakhot_probability_1989,procaccia_transitions_1991}. By using the TC-RB analogy \citep{eckhardt_fluxes_2007}, the skewness of the PDFs is similarly induced by plumes of angular velocity, with higher speeds dominating near the inner cylinder where herringbone-like patterns of streaks can form \citep{dong_direct_2007,froitzheim_statistics_2019}. The left and right tails of the PDFs are associated with low-speed and high-speed fluid, respectively. The high-speed fluid experiences greater centrifugal force that cannot be balanced by the pressure gradient in the radial direction. Consequently, it will be ejected from the velocity boundary layer, and this fluid is referred to as a plume. Therefore, the shrinkage of the right tails with the increase in the volume fraction $\phi$ may indicate the suppression of angular velocity plumes.

To verify our conjecture, we show the contours of the azimuthal velocity from our numerical simulation in Fig. \ref{fig:pdf&simulation}. The simulation details of the turbulent emulsion in TC flow can be found in our previous studies \citep{su_turbulence_2024,su_numerical_2024}. For the two-phase case with $\phi=20\%$, we also plot the interfaces of the droplets, shown as solid lines in Fig. \ref{fig:pdf&simulation}(\textit{c}). In the single-phase case (Fig. \ref{fig:pdf&simulation}(\textit{b})), the plumes prefer to eject from the outflow region due to the presence of the Taylor vortex \citep{dong_direct_2007}. In the two-phase case (Fig. \ref{fig:pdf&simulation}(\textit{c})), the droplet interface acts as a wall, preventing the plumes from penetrating through the interface. In this way, droplets block the ejection of plumes from the inner cylinder. This process results in the shrinkage of the PDFs in their right tails. As a result, both the mean velocity and rms velocity fluctuation are reduced, as shown in Fig. \ref{fig:IW-fluc}(\textit{b,c}).

{
When considering the global torque, it is directly related to the angular velocity flux $J_\omega$, which is conserved in the radial direction, by the relation $J_\omega = G\nu_c^2$ \citep{eckhardt_fluxes_2007,eckhardt_torque_2007}. $J_\omega$ is defined as
\begin{equation}
	J_\omega = J_{\omega,adv} + J_{\omega,dif} + J_{\omega,int},
	\label{equ:J_w}
\end{equation}
where the three terms represent \citep{hori_interfacial-dominated_2023}
\begin{enumerate}
	\item the advective contribution $J_{\omega,adv} = r^3\langle u_r \omega \rangle_{t,z}$ where $\omega = u_\theta / r$ is the angular velocity;
	\item the diffusive contribution $J_{\omega,dif} = -r^3\nu \partial_r \langle \omega \rangle_{t,z}$;
	\item the interfacial contribution $J_{\omega,int}$.
\end{enumerate}
The interfacial contribution cannot be measured in experiment and we refer to our recent work \citep{Su_Zhang_Wang_Yi_Xu_Fan_Wang_Sun_2025} for more details. For single-phase laminar flow, the first and third term vanishes due to $u_r=0$, so only the second term contributes \citep{eckhardt_torque_2007}: 
\begin{equation}
	J_{\omega,lam}=2\nu r_i^2 r_o^2 \omega_i /(r_o^2-r_i^2).
	\label{equ:J_w_lam}
\end{equation}
For two-phase flow, near the cylinder surface, the diffusive term plays a dominant role due to the no-slip condition and high angular velocity gradient. Consequently, the first term $J_{\omega,adv}$ in Eq. \ref{equ:J_w} can be neglected, and the second term $J_{\omega,dif}$ will dominate near the inner cylinder \citep{wang_numerical_2023,su_numerical_2024,su_turbulence_2024}. The decrease of the mean azimuthal velocity $\langle u_\theta \rangle_{t,z}$ in Fig. \ref{fig:IW-fluc}(\textit{b}) implies that the magnitude of the wall-normal velocity gradient $\partial_r \langle \omega \rangle_{t,z}$ will be augmented for a fixed $\widetilde{r}$. Therefore, the angular velocity flux $J_\omega$ as well as the global torque $G$ are enhanced. 
}



\subsection{Modulation of continuous phase at the middle gap}
\label{subsec:M}

\begin{figure}
	\centerline{\includegraphics[width=0.95\textwidth]{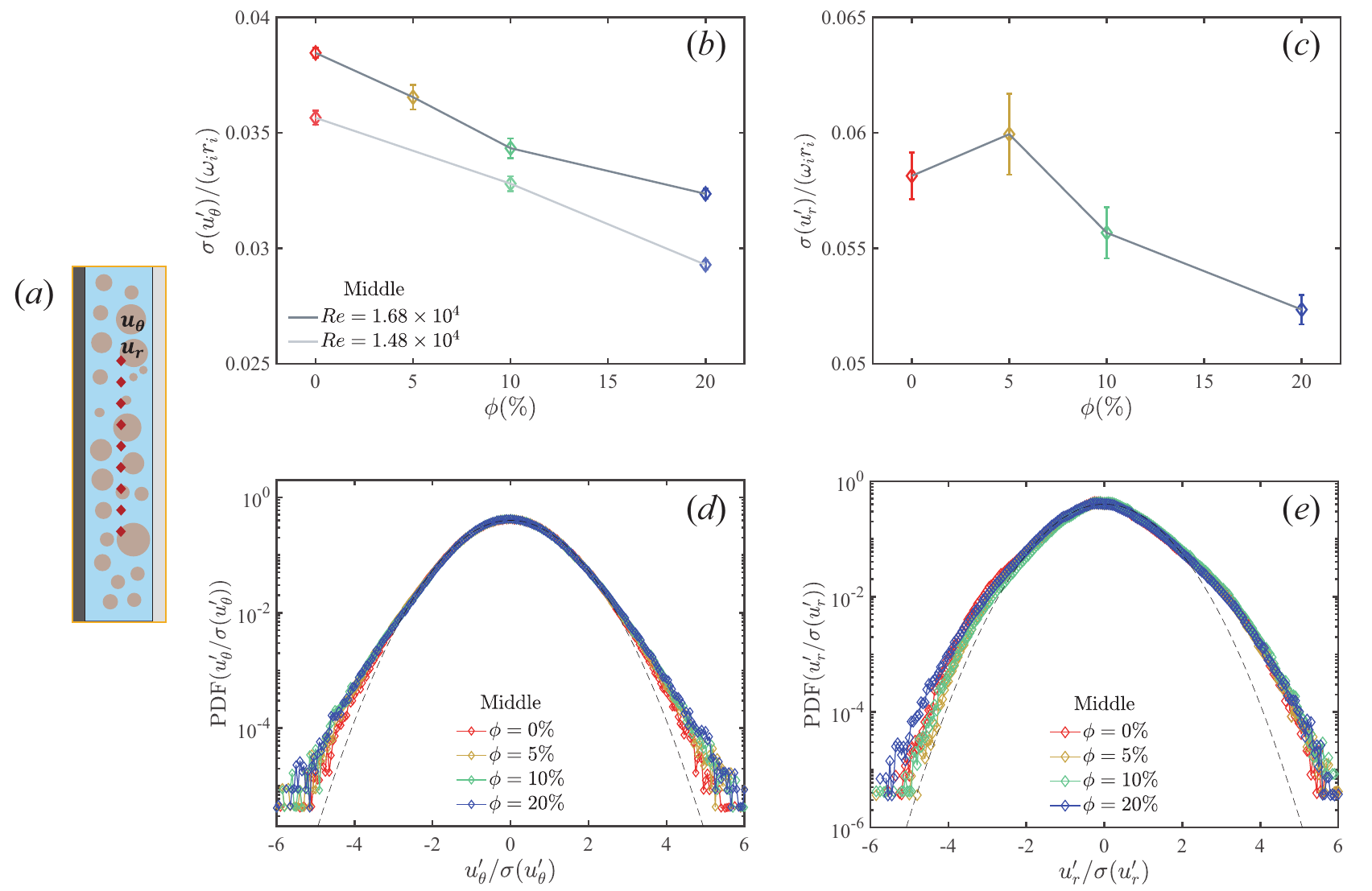}}
	\caption{(\textit{a}) A schematic shows the LDA measurement position at the midgap, represented by the red diamonds. Both azimuthal velocity $u_\theta$ and radial velocity $u_r$ can be measured here. The root mean squared velocity fluctuations for the azimuthal (\textit{b}) and radial (\textit{c}) velocities are shown with varying volume fraction $\phi$ for two different Reynolds numbers. Probability density functions of the normalized azimuthal (\textit{d}) and radial (\textit{e}) velocities are also presented. The gray dashed lines correspond to the standard normal distribution.}
	\label{fig:mid-fluc}
\end{figure}

We then investigate the turbulence statistics at the middle gap, where both the azimuthal velocity $u_\theta$ and radial velocity $u_r$ are measured (see Fig. \ref{fig:mid-fluc}(\textit{a})). The rms velocity fluctuations $\sigma(u_\theta^\prime)$ and $\sigma(u_r^\prime)$ are shown in Fig. \ref{fig:mid-fluc}(\textit{b,c}). The velocity fluctuations decrease with the increase in volume fraction. {Note that the small peak in Fig. \ref{fig:mid-fluc}(\textit{c}) at $\phi = 5\%$ is within the error bar of our experimental measurement because $u_r$ is much smaller than $u_\theta$ and therefore has a large uncertainty.} In the bulk region of TC turbulence, the flow is close to homogeneous and isotropic turbulence \citep{grossmann_highreynolds_2016}. The suppression of velocity fluctuations by dispersed droplets has also been observed in numerical simulations of homogeneous and isotropic turbulence \citep{dodd_interaction_2016,mukherjee_dropletturbulence_2019,crialesi-esposito_modulation_2022}, where the suppression effect is attributed to the break-up of droplets by the large-scale turbulent fluctuations. During the break-up process, the turbulent kinetic energy is transferred to the interfacial energy stored by the droplets.

At first glance, our results seem to be consistent with the numerical simulations of homogeneous and isotropic turbulence \citep{dodd_interaction_2016,mukherjee_dropletturbulence_2019,crialesi-esposito_modulation_2022}. However, this break-up mechanism cannot be applied to the middle gap of the TC turbulence. In our previous study, we found that the droplets cannot be fragmented by the turbulence fluctuations in the bulk region \citep{yi_physical_2022}. To provide further evidence for this conjecture, we estimate the local Weber number using the measured velocity information. The local Weber number is defined as
\begin{equation}
	We = \frac{\rho_c \langle {\delta u}_D ^2 \rangle_{t,z} \langle D \rangle}{\gamma}
	\label{equ:We},
\end{equation}
where $\langle {\delta u}_D ^2 \rangle_{t,z}$ is the average squared velocity increment over a distance equal to the droplet mean diameter $\langle D \rangle$ \citep{risso1998oscillations}. In our experiment, $\langle {\delta u}_D ^2 \rangle_{t,z}$ can be estimated as follow
\begin{equation}
	\langle {\delta u}_D ^2 \rangle_{t,z}= \langle {\delta u^2_\theta}(\tau=\langle D \rangle/ \langle u_\theta \rangle_{t,z}) \rangle_{t,z}
	\label{equ:du},
\end{equation}
where $\delta u_\theta(\tau) = u_\theta(t+\tau) - u_\theta(t)$ is the velocity increment over a time interval $\tau$, and $\tau = \langle D \rangle / \langle u_\theta \rangle_{t,z}$ is the average time interval corresponding to the mean droplet diameter. Here we have invoked the Taylor frozen hypothesis since the turbulence fluctuation is much smaller than the mean velocity \citep{huisman_statistics_2013}. Besides, $\langle \delta u_D^2 \rangle_{t,z}$ is calculated from the single phase case, considering that the droplet size is measured at a volume fraction of $\phi=1\%$. The value of local Weber number $We$ at the middle gap is {$We_{\Tilde{r}=0.5} \approx 0.17 < 1$}, suggesting that the turbulent fluctuation is not strong enough to break up the droplets in the bulk region \citep{yi_physical_2022}. {At higher volume fractions, the Weber number is $We_{\Tilde{r}=0.5} \approx 0.14, 0.12$ and $0.11$ at $\phi = 5\%, 10\%$ and $20\%$, respectively, based on the same mean droplet diameter.} Thus, the suppression of fluctuation in the bulk cannot be attributed to break-up induced energy transfer. The suppression of the velocity fluctuation in the bulk can be understood as follows. The frequent burst of plumes also transfers energy from the inner cylinder to the bulk region. {Since droplet interfaces block the emission of velocity plumes, less energy is injected into the bulk region resulting in lower turbulent fluctuations in both the azimuthal and radial directions}.



We also include the PDFs of $u_\theta ' /\sigma(u_\theta ')$ and $u_r '/\sigma(u_r ') $ in Fig. \ref{fig:mid-fluc}(\textit{d,e}). $u_\theta '$ and $u_r '$ are normalized by their rms velocity fluctuation. All PDFs are close to Gaussian distribution represented by the dashed lines, in line with previous study \citep{froitzheim_statistics_2019}. After normalization, the PDFs collapse in their middle part. However, the tails of PDFs tend to become fatter as the volume fraction $\phi$ increases, indicating that turbulent emulsions are more intermittent than single phase ones at large scales{, i.e. the intermittency of velocity fluctuation is enhanced.} This behaviour has also been reported in droplet-laden homogeneous isotropic turbulence \citep{crialesi-esposito_modulation_2022}. {The enhanced intermittency in the bulk region at large scales originates from the presence of droplets. When the droplets deform and bounce back to the original spherical shape, the continuous phase adjacent to the droplets will experience extreme events due to the interfacial tension. This results in the intermittency enhancement at large scales at the middle gap. }

\begin{figure}
	\centerline{\includegraphics[width=0.95\textwidth]{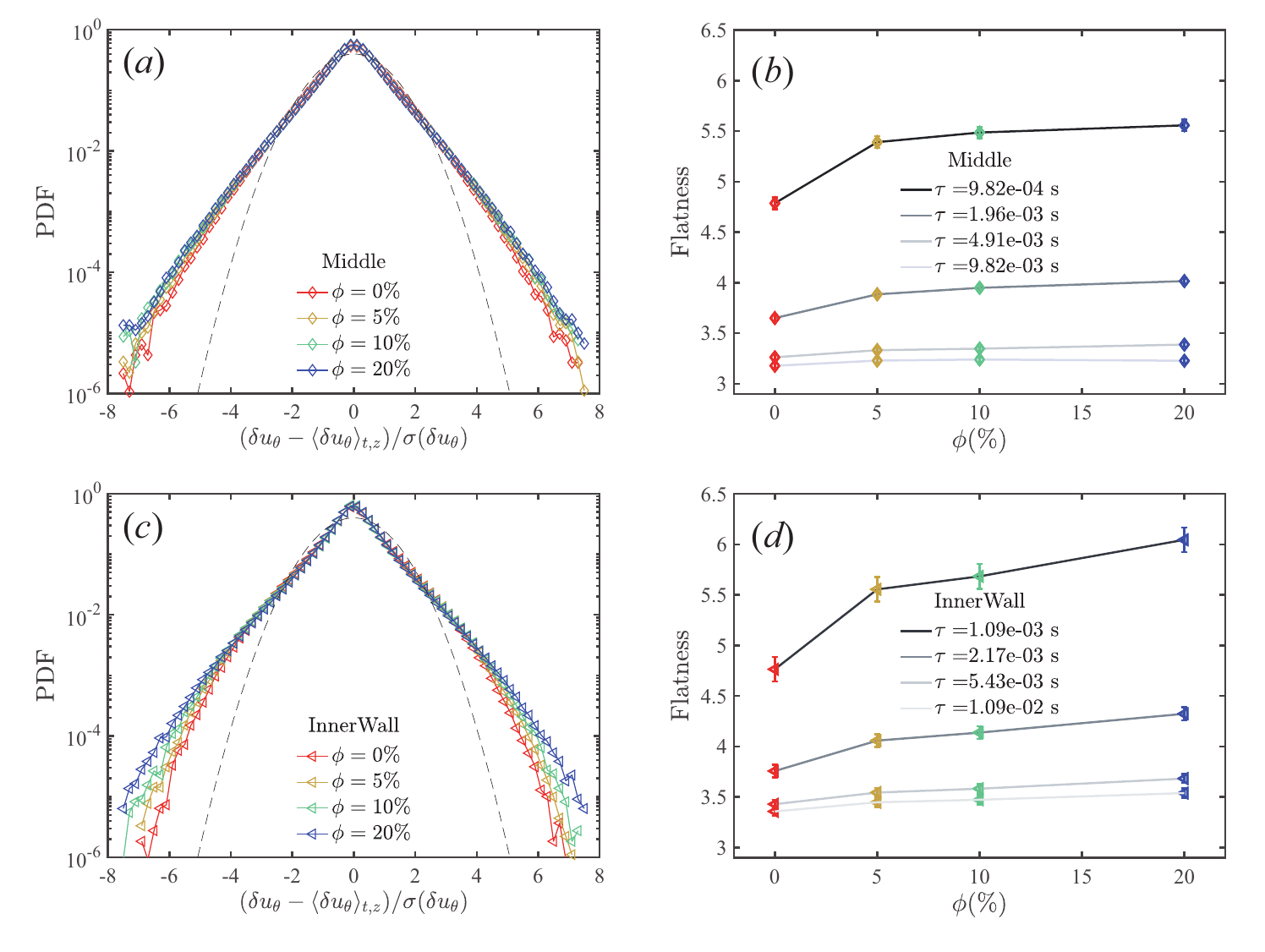}}
	\caption{Probability density functions of the velocity increment of azimuthal velocity component normalized by its mean and standard deviation at the middle gap (\textit{a}) and near the inner cylinder (\textit{c}). Here, the time interval $\tau=9.82\times 10^{-4}$ s in (\textit{a}) and $\tau=1.09\times 10^{-3}$ s in {(\textit{c})}. The gray dashed lines refer to the standard normal distribution. Flatness of velocity increment for several typical values of time interval with varying volume fraction at the middle gap (\textit{b}) and near the inner cylinder (\textit{d}).}
	\label{fig:increment}
\end{figure}

Small-scale intermittency can be probed using the velocity increment $\delta u_\theta(\tau)$. The normalized PDFs of $\delta u_\theta$, scaled by their mean and standard deviation, are plotted in Fig. \ref{fig:increment}(\textit{a}) for the smallest time interval, $\tau=9.82\times 10^{-4}$ s, obtained in our experiment. The PDFs of $\delta u_\theta$ for a small $\tau$ are non-Gaussian, indicating small-scale intermittency. As the volume fraction increases, the tails of the PDFs become heavier than those in the single-phase case. To quantify this, we calculate the flatness of $\delta u_\theta(\tau)$, $\langle \delta u_\theta^4 (\tau) \rangle_{t,z}/\langle \delta u_\theta^2 (\tau) \rangle^2_{t,z}$, for several values of $\tau$, shown in Fig. \ref{fig:increment}(\textit{b}). 
Taking the gap width $d$ as the integral length scale and {the azimuthal velocity of inner cylinder as the characteristic velocity}, the corresponding integral time scale 
{$t_i=d/r_i \omega_i \approx 4.7\times 10^{-3} $ s, 
at which the level of intermittency should be similar to what we reported in Fig. \ref{fig:mid-fluc}(\textit{d,e}).} In general, the flatness of $\delta u_\theta(\tau)$ increases with the volume fraction of the dispersed phase for $\tau$ ranging from the smallest value measured in our experiment to the {twice} integral time scale. Similarly, we present the PDFs of the normalized $\delta u_\theta$ and their flatness for data measured near the inner cylinder in Fig. \ref{fig:increment}(\textit{c,d}). Near the inner cylinder, the change in the PDF tails becomes more pronounced compared to the bulk region. The flatness also increases monotonically with the volume fraction. These results indicate that intermittency across the entire range of scales resolved here is enhanced in both the near-wall and bulk regions. We also note that the enhancement of intermittency is stronger in the near-wall region than in the bulk.

Intermittency in multiphase turbulence has attracted growing interest in recent years, driven by advancements in numerical simulations. \cite{crialesi-esposito_modulation_2022,crialesi-esposito_intermittency_2023} investigated the PDFs of velocity and velocity increments, $\delta u$, between points conditioned to be located in the same phase or different phases. They found that the leading contribution to the increased deviation from Gaussian statistics at small scales comes from velocity increments across phase interfaces. However, the PDFs of $\delta u/\sigma(\delta u)_{SP}$, conditioned on points belonging to the continuous phase, remained nearly unchanged, where $\sigma(\delta u)_{SP}$ is the standard deviation of $\delta u$ in the single-phase case. In our study, we use $\sigma(\delta u_\theta)$ from each respective case to normalize the velocity increment, as the flatness measures deviations from $\sigma(\delta u_\theta)$, not $\sigma(\delta u_\theta)_{SP}$. \cite{crialesi-esposito_how_2024} further showed that the breakup of large droplets and the rupture of ligaments generate high vorticity and strain, which increases small-scale intermittency. 

In light of these findings in homogeneous isotropic turbulence, we can interpret our results in TC turbulence. In the bulk region {where the flow is nearly homogeneous and isotropic}, the turbulence fluctuation is not strong enough to break up the droplets, as we have previously shown. The increased intermittency originates from the presence of interfaces {as found by \citep{crialesi-esposito_intermittency_2023}}. {Near the inner cylinder, droplets can be fragmented by the higher levels of shear, i.e., the dynamic pressure induced by the large mean velocity gradient \citep{levich1962physicochemical,yi_physical_2022}. We also note that the velocity plumes are able to deform large droplets, as shown in Fig. \ref{fig:pdf&simulation}(\textit{c}). During this process, the plume loses its energy in favor of the interface deformation. The large droplet deformation should then lead to the formation of several small scales structures, hence being responsible for the increase of intermittency. Therefore, the intermittency enhancement near the inner cylinder originates from not only the presence of interfaces but also the fragmentation and deformation of droplets. Consequently, additional intermittency beyond the presence of interfaces can be observed near the inner cylinder. }

As we mentioned before, the global torque is directly related to angular velocity flux $J_\omega$ defined in Eq. \ref{equ:J_w}. At the middle gap, the second term could be neglected due to the fact that the angular velocity gradient is small throughout the bulk region \citep{dong_direct_2007,froitzheim_statistics_2019,su_numerical_2024}. In TC flow, the turbulence is a combination of turbulent Taylor vortices and background fluctuations. {Thus, the first term $J_{\omega,adv} = r^3\langle u_r \omega \rangle_{t,z}$ can be further decomposed into two parts \citep{brauckmann_direct_2013}}:
\begin{equation}
	r^3\langle u_r \omega \rangle_{t,z} = r^3\langle \langle u_r \rangle_{t} \langle \omega \rangle_{t} \rangle_z + r^3\langle u_r ' \omega ' \rangle_{t,z} . 
	\label{equ:decomposition}
\end{equation} 
These two terms are contributed by mean Taylor vortices and the turbulent fluctuation motion and can be denoted by their dimensionless forms, {$Nu^M = r^3\langle \langle u_r \rangle_{t} \langle \omega \rangle_{t} \rangle_z /J_{\omega,lam}$} and $Nu^T = r^3\langle u_r ' \omega ' \rangle_{t,z}/J_{\omega,lam}=r^2 \langle u_\theta ' u_r ' \rangle_{t,z}/J_{\omega,lam}$, respectively.

\begin{figure}
	\centerline{\includegraphics[width=0.95\textwidth]{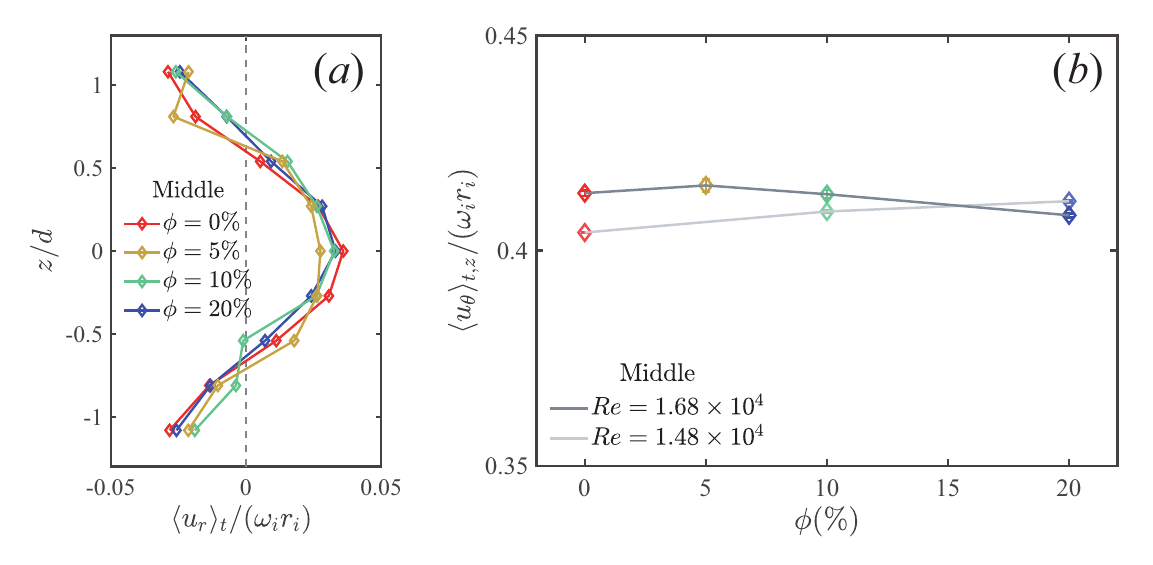}}
	\caption{(\textit{a}) {Axial profiles} of the mean radial velocity, normalized by the inner cylinder velocity, $\langle u_r \rangle_t/(\omega_i r_i)$, {for different volume fractions from $\phi= 0\%$ to $\phi=20\%$ at $Re=1.68 \times 10^4$.}  (\textit{b}) Mean azimuthal velocity, normalized by the inner cylinder velocity, $\langle u_\theta \rangle_{t,z}/(\omega_i r_i)$, with varying volume fraction $\phi$ for two Reynolds numbers.}
	\label{fig:mid-mean}
\end{figure}

We first study the effect of droplets on the Taylor vortices. We plot in {Fig. \ref{fig:mid-mean}(\textit{a})} the axial profiles of the radial velocity for {different volume fractions from $\phi = 0\%$ to $\phi = 20\%$ }
at $Re=1.68 \times 10^4$ 
and the average azimuthal velocity with varying $\phi$ at two different $\Rey$ in Fig. \ref{fig:mid-mean}(\textit{b}). These $u_r$ profiles approximately overlap with each other. For $\langle u_\theta \rangle_{t,z}$, it also remains nearly unchanged at various $\phi$ explored in our experiment, implying that the existence of droplets would not change the mean flow dramatically, at least in the present parameter range. In all, the angular velocity flux and global torque contributed by the Taylor vortices in the bulk region remain almost unchanged as the droplets have a marginal effect on them.

\begin{figure}
    \centerline{\includegraphics[width=0.95\textwidth]{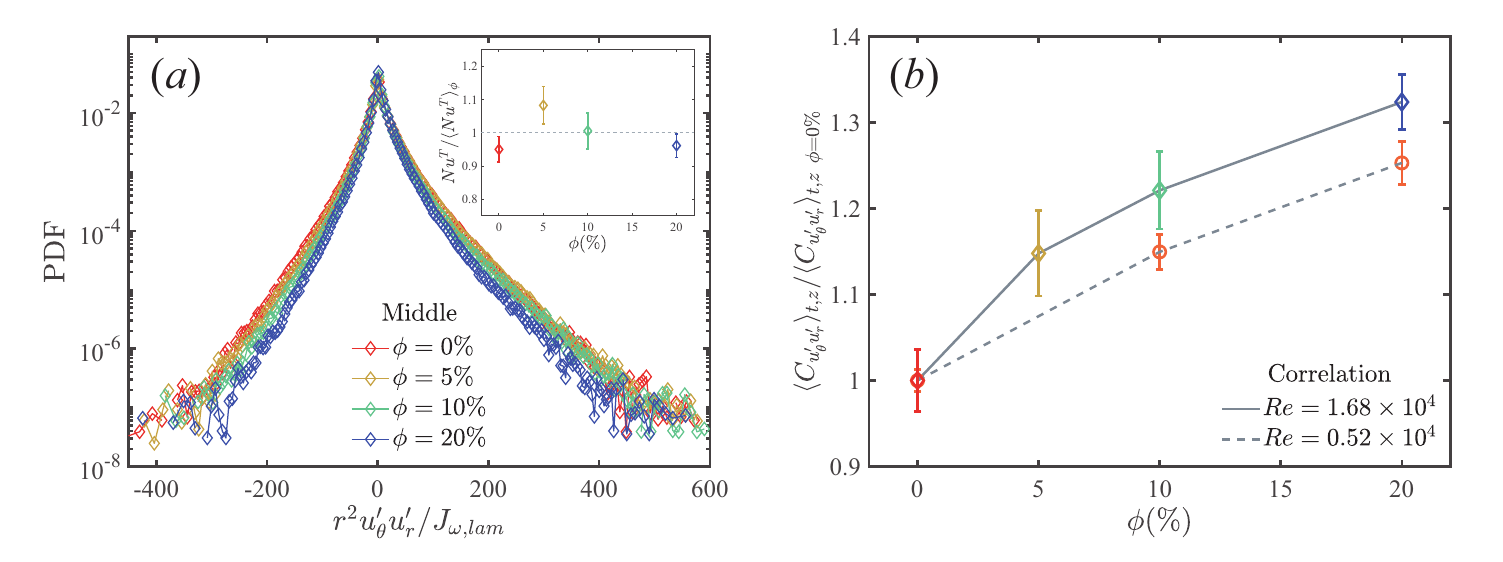}}
    \caption{ (\textit{a}) Probability density functions of dimensionless angular velocity flux normalized by the corresponding laminar angular velocity flux, i.e., $r^2 u_\theta ' u_r '/(J_{\omega,lam})$. {Inset: average angular velocity flux contributed by the turbulent flow $Nu^T$ normalized by $\langle Nu^T \rangle_{\phi}$, which is the average Nusselt number over all volume fractions.} Data are from $Re=1.68 \times 10^4$. (\textit{b}) The cross correlation coefficients between $u_\theta '$ and $u_r '$ as a function of volume fraction normalized by the single phase case at $Re=1.68 \times 10^4$ (diamond mark) measured in our experiment and at $\Rey = 0.52 \times 10^4$ obtained from the numerical simulation (circle mark) \citep{su_turbulence_2024}.}
    \label{fig:Nu}
\end{figure}

We then explore the turbulence contribution to the angular velocity flux and plot PDFs of dimensionless angular velocity flux $r^2 u_\theta ' u_r ' / J_{\omega, lam}$ in Fig. \ref{fig:Nu}(\textit{a}). {Note that when calculating the production of $u_\theta '$ and $u_r '$, the series of velocity component with higher data rate is interpolated based on the other one with lower data rate.} The PDFs are all positively skewed since the momentum has to be transported to the outer cylinder, which is consistent with previous studies \citep{huisman_ultimate_2012,brauckmann_direct_2013,froitzheim_statistics_2019}. Furthermore, they are highly non-Gaussian, which is reflected by the observation that the local flux $r^2 u_\theta ' u_r '$ taking value $\pm 400$ times as large as $J_{\omega,lam}$ still has a high probability of occurrence \citep{huisman_ultimate_2012, froitzheim_statistics_2019}. These large rare events are found to be footprints of angular velocity flux fluctuations induced by plumes, similar behaviours are also reported in Rayleigh-B\'enard turbulence \citep{shang2003measured}. With the presence of droplets, both tails of the PDFs shrink as $\phi$ increases, implying that the large rare events are suppressed. The reduced probability of rare angular velocity flux events is an additional sign that the droplet interfaces can block the emission of angular velocity plumes near the inner cylinder, providing further evidence to our conclusion made in \S \ref{subsec:IW}. {We also note that these large rare events make the measurement of $Nu^T$ quite challenging as already reported in \cite{huisman_ultimate_2012} and \cite{froitzheim2017velocity}, which results in uncertainty in $Nu^T$.}

Since the droplets inhibit the turbulent fluctuations of both the azimuthal and radial velocity (the percentage of depression is up to 15\% for $\sigma(u^\prime_\theta)$ at $\phi=20\%$), we would expect that the angular velocity flux contributed from the turbulence would be suppressed accordingly, i.e., the overall profiles of PDFs for two-phase cases would be shifted horizontally. However, the decreasing tendency of $Nu^T$ is not observed in our experiment. Instead, $Nu^T$ fluctuates around a constant within the experimental uncertainties (see the inset of Fig. \ref{fig:Nu}(\textit{a})). Thus, the angular velocity flux and global torque contributed by the turbulent flow in the bulk region also remain almost unchanged.

Taken together, we show that the advective contributions to global torque by the mean flow and turbulence fluctuations remain nearly unmodified in the bulk region. 
We therefore conclude that the drag enhancement presumably originates from the effect of two-phase interfaces, which cannot be quantified in the experiment but has already been investigated in our previous simulation work \citep{su_numerical_2024,su_turbulence_2024}.

From the inset of Fig. \ref{fig:Nu}(\textit{a}), {we found that $Nu^T/\langle Nu^T \rangle_{\phi}$ fluctuates  around 1, and does not show a $\phi$ dependence.} To investigate the reason of nearly constant $Nu^T$, we make a further conversion of the second term in Eq. \ref{equ:decomposition}: 
\begin{equation}
	r^3\langle u_r ' \omega ' \rangle_{t,z} = r^2 \langle u_r ' u_\theta ' \rangle_{t,z} = r^2 \sigma(u_\theta ') \sigma(u_r ') \langle C_{u_\theta ' u_r '} \rangle_{t,z},
	\label{equ:J_w_turb_correlation}
\end{equation}
where $\langle C_{u_\theta ' u_r '} \rangle_{t,z}$ refers to the cross correlation coefficient between the azimuthal and radial velocity, which is defined as
\begin{equation}
	\langle C_{u_\theta ' u_r '} \rangle_{t,z} = \frac{\langle u_\theta ' u_r ' \rangle_{t,z}}{\sigma(u_\theta ') \sigma(u_r ')} .
	\label{equ:correlation}
\end{equation}
{The concept of cross-correlation has been widely adopted in the single-phase TC turbulence \citep{burin_local_2010,brauckmann_momentum_2016,huisman_ultimate_2012}. Eq. \ref{equ:J_w_turb_correlation} indicates that only the correlated fluctuations of $u_\theta '$ and $u_r '$ contribute to the net convective transportation.} 
Therefore, the cross correlation coefficient is calculated for varying $\phi$ and averaged over different heights, being illustrated in Fig. \ref{fig:Nu}(\textit{b}), where we also include the results from our numerical simulation. For the two phase case, $\langle C_{u_\theta ' u_r '} \rangle_{t,z}$ exhibits clear and monotonic enhancement behaviours with the increase of $\phi$. The enhancement of $\langle C_{u_\theta ' u_r '} \rangle_{t,z}$ is more than 30\% at $\phi=20 \%$, indicating that the turbulent emulsion becomes more coherent than the single phase flow due to the presence of droplets.  $\langle C_{u_\theta ' u_r '} \rangle_{t,z}$ from the numerical simulation also displays an enhancement but to a less degree, which may be related to the different Reynolds number in the experiments and numerical simulation. {Thus, the nearly constant of $Nu^T$ can be attributed to the enhanced coherence.} Since the turbulence becomes more intermittent as discussed above and the breakup and coalescence events of droplets are found to create disturbances on coherent structures \citep{dodd_interaction_2016,crialesi-esposito_modulation_2022,perlekar2019kinetic}, we are surprised to observe that the turbulence coherence is enhanced by the dispersed droplets. In the case of turbulent emulsion, the continuous phase is surrounded by dispersed droplets. The continuous phase fluid thus cannot evolve freely like it  could do in the single phase case due to the confinement of the surrounding droplets. This conjecture needs to be tested in future studies.



\subsection{Velocity power spectrum}
\label{subsec:spectrum}

\begin{figure}
	\centerline{\includegraphics[width=0.99\textwidth]{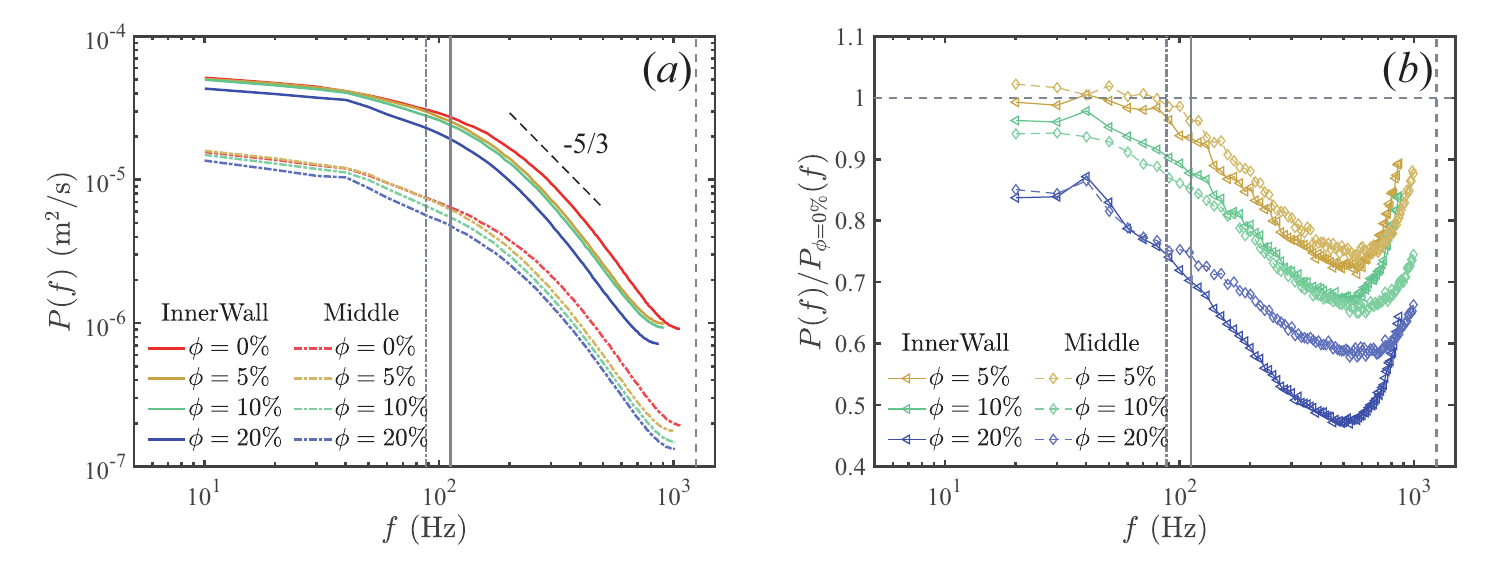}}
	\caption{(\textit{a}) Velocity power spectra $P(f)$ for the azimuthal velocity $u_\theta '$ near the inner cylinder (solid line --) and at the middle gap (dash-dotted line -- $\cdot$). The thin dashed line with a slope of $-5/3$ represents the inertial range scaling exponent at high Reynolds numbers. (\textit{b}) $P(f)$ for the two-phase cases is scaled by its single-phase counterpart, i.e., $P(f)/P_{\phi=0\%}(f)$. The vertical gray solid and dash-dotted lines refer to the characteristic frequencies corresponding to the gap width, $\langle u_\theta \rangle_{t,z} / d$, near the inner cylinder and at the middle gap, respectively. The vertical gray dashed line denotes the frequency, $\langle u_\theta \rangle_{t,z} / \langle D \rangle$, corresponding to the mean droplet diameter at the middle gap, where we conduct the droplet size measurement. Data are from $Re=1.68 \times 10^4$.}
	\label{fig:spectrum}
\end{figure}

As we have discussed, the turbulence fluctuation is suppressed by the presence of droplets near the inner cylinder and at the middle gap. It is known that turbulence implies fluid motion across a wide spectrum of length and time scales \citep{Pope_2000}. Thus, it is also instructive to investigate the effect of droplets on the multi-scale nature of turbulence. This can be accomplished by inspecting the velocity power spectra of the azimuthal velocity fluctuation $u_\theta '$. Because the arrival times of LDA measurements are stochastic in nature, the time series are then linearly interpolated using twice the average acquisition frequency \citep{huisman_statistics_2013}, aiming to create a time series with equal temporal spacing, which facilitates the application of fast Fourier transformation. The power spectra near the inner cylinder, as well as at the middle gap for different volume fractions, are shown in Fig. \ref{fig:spectrum}(\textit{a}). We also illustrate two characteristic frequencies, namely the frequency corresponding to the gap width $\langle u_\theta \rangle_{t,z}/d$ near the inner cylinder (vertical solid line) and at the middle gap (dashed-dotted line), and the frequency corresponding to the mean droplet diameter $\langle u_\theta \rangle_{t,z}/\langle D \rangle$ (vertical dashed line). Due to measurement limitations, frequencies higher than $\langle u_\theta \rangle_{t,z}/\langle D \rangle$ cannot be resolved in this study. The inertial range scaling is expected to appear in the range of frequencies $f>\langle u_\theta \rangle_{t,z}/d$. In the bulk, we also use the Taylor scale-based Reynolds number to measure turbulence intensity, which is defined as
\begin{equation}
    R_\lambda = \frac{\sigma(u_\theta ') \lambda}{\nu_c} = \sqrt{\frac{15 \sigma(u_\theta ')^4 \epsilon}{\nu_c}},
\end{equation} 
where the average energy dissipation rate $\epsilon$ is estimated by the global torque $\epsilon = 0.1T \omega_i/(\pi (r_o^2-r_i^2)L\rho_c)$ \citep{ezeta2018turbulence,yi_physical_2022}. At $Re=1.68 \times 10^4$, $R_\lambda \approx 70$. The $R_\lambda$ is too low to observe a clear inertial range scaling. Nevertheless, we show the $-5/3$ scaling (thin dashed line) in Fig. \ref{fig:spectrum}(\textit{a}) as a reference.



Firstly, the amplitude of the velocity power spectra obtained near the inner cylinder is much higher than that at the middle gap, indicating that the turbulence near the inner cylinder is substantially stronger than at the middle gap. This finding was also captured by \citet{dong_direct_2007}, which shows an uneven distribution and asymmetry in the intensity of turbulent fluctuations caused by the curvature effect. When comparing the single-phase and two-phase cases near the inner cylinder, {it seems that at very large scales the kinetic energy increases for $\phi = 5\%$ case. However, the deviation of the $\phi=5\%$ case from 1 is less than $3\%$ which could be seen from Fig. \ref{fig:spectrum}(\textit{b}). This deviation is small and within the experimental uncertainty. Therefore, it can be concluded that} the energy content of turbulent fluctuations at all scales resolved in our experiment is strongly depressed, especially at scales smaller than $d$, i.e., frequencies higher than $\langle u_\theta \rangle_{t,z}/d$. This result is consistent with previous simulation works \citep{perlekar_spinodal_2014,dodd_interaction_2016,mukherjee_dropletturbulence_2019,crialesi-esposito_modulation_2022,crialesi-esposito_intermittency_2023,trefftz-posada_interaction_2023,rosti_droplets_2019}. Even though these simulations studied different kinds of turbulent flows, such as homogeneous isotropic turbulence \citep{perlekar_spinodal_2014,dodd_interaction_2016,mukherjee_dropletturbulence_2019,crialesi-esposito_modulation_2022,crialesi-esposito_interaction_2023} and homogeneous shear turbulence \citep{rosti_droplets_2019,trefftz-posada_interaction_2023}, energy spectrum suppression at scales larger than the average droplet diameter is consistently observed. Moreover, as the number of droplets increases, the degree of suppression on turbulence fluctuations also rises.

Regarding the mechanism, one widely accepted hypothesis is as follows: In Newtonian turbulence, the non-linear energy flux of the fluid is the only energy transfer mechanism. In turbulent emulsions, droplets provide an additional mechanism of energy transfer due to the frequent breakup and deformation of the droplets; the turbulent kinetic energy can be transferred from the fluid to the interfacial energy stored by the droplets \citep{perlekar2019kinetic,crialesi-esposito_modulation_2022}. The higher the volume fraction, the greater the likelihood of deformation and breakup events, resulting in a higher degree of suppression of turbulent kinetic energy at scales larger than the mean droplet size. This hypothesis offers a reasonable explanation for the energy spectrum suppression near the inner cylinder, while at the middle gap, the suppression results from reduced ejection of plumes near the inner cylinder, leading to less energy being transported into the bulk region. The simulation results also indicate that the spectrum at scales smaller than the mean droplet diameter is greatly enhanced, presumably due to the coalescence of small droplets, which feeds the interfacial energy back into the fluid. The increased energy content at scales smaller than the mean droplet diameter needs to be verified in future studies.

To compare the effects of droplets on the spectrum near the inner cylinder and at the middle gap, the spectrum of the two-phase case is scaled by its single-phase counterpart, i.e., $P(f)/P_{\phi=0\%}(f)$, as shown in Fig. \ref{fig:spectrum}(\textit{b}). It is clear that the suppression effect becomes stronger as the frequency increases (or the scale decreases). The faster decay of $P(f)$ when $f > \langle u_\theta \rangle_{t,z}/d$ suggests that the scaling of the spectrum is modified, as also reported in \cite{perlekar_spinodal_2014}. We note that the roll-up of $P(f)/P_{\phi=0\%}(f)$ at high frequency is due to instrument noise \citep{huisman_statistics_2013}, which is also reflected in the leveling-off of $P(f)$ in Fig. \ref{fig:spectrum}(\textit{a}).

At low volume fractions ($\phi = 5\%$ and $10\%$), the $P(f)/P_{\phi=0\%}(f)$ curves nearly collapse near the inner cylinder and at the middle gap, indicating that the suppression effect is similar at these two positions. 
At a higher volume fraction ($\phi = 20\%$), the suppression effect is stronger near the inner cylinder than at the middle gap for frequencies higher than $\langle u_\theta \rangle_{t,z}/d$ (scales smaller than the gap width $d$). One possible explanation is that, at high volume fractions, the droplets have a higher probability of merging with each other, which in turn releases the interfacial energy back to the fluid at the middle gap since the droplets are less likely to break apart there.



\section{Conclusion}
\label{sec:conclusion}
In this work, we experimentally investigate how the dispersed droplets modulate the global drag and the statistical properties of turbulence velocity fluctuations in the continuous phase of a turbulent emulsion. The emulsion is generated in a turbulent Taylor-Couette flow. We precisely match the refractive indices of the dispersed and continuous phases, which facilitates our measurement of the local velocity of the continuous phase for volume fractions of up to $20\%$. Due to the inhomogeneity of the Taylor-Couette flow, the velocity measurements are performed at two representative radial locations: near the inner cylinder and at the middle gap.

Near the inner cylinder, the droplet interfaces can suppress the emission of angular velocity plumes from the boundary layer, resulting in a reduction in both the mean azimuthal velocity and the root mean squared fluctuation of azimuthal velocity. This reduction in the mean azimuthal velocity leads to a higher gradient of angular velocity in the radial direction, thus increasing the viscous diffusion contribution to the angular velocity flux and global drag of the system. By comparing the velocity power spectrum, we find that the energy content at scales above the average droplet diameter is depressed, which results from the reduced emission of plumes and breakup-induced energy transfer from the fluid to the interfacial energy of the droplets.

At the middle gap, the angular velocity flux contributed by the mean Taylor vortex, $Nu^M$, and the turbulence, $Nu^T$, are nearly unaltered by the droplets. The droplets enhance the cross-correlation between the angular and radial velocity, which compensates for the reduction in velocity fluctuation, leaving $Nu^T$ unchanged. However, the rare events of angular velocity flux contributed by the turbulent flow are less frequent due to the reduced plume emission. We further show that the intermittency of velocity increments is enhanced due to the presence of interfaces. A similar behavior is observed near the inner cylinder, but the degree of enhancement is higher at small scales, which may be related to the breakup-induced generation of high vorticity and strain.

In this study, we present the first attempt to gather velocity information in a wall-bounded turbulent emulsion. Our study provides valuable experimental insights into how the dispersed droplets modulate global drag, coherent structures, and the multi-scale nature of the turbulent flow. In future work, we aim to further improve the measurement techniques to simultaneously capture the velocities of both the continuous and dispersed phases in dense turbulent emulsions, studying the detailed coupling dynamics of the two phases.

\backsection[Acknowledgements]{We acknowledge Yingzheng Liu for insightful discussions on the experimental methods. We thank Federico Toschi, Xander M. de Wit, and Sander Huisman for their insightful discussions and suggestions.}

\backsection[Funding]{This work is financially supported by the National Natural Science Foundation of China under Grant Nos. 11988102, 12402298 and 12402299, the New Cornerstone Science Foundation through the New Cornerstone Investigator Program and the XPLORER PRIZE.}

\backsection[Declaration of interests]{The authors report no conflict of interest.}


\backsection[Author ORCIDs]{\\
	Yaning Fan https://orcid.org/0009-0000-5886-3544\\
	Yi-Bao Zhang https://orcid.org/0000-0002-4819-0558; \\
	Jinghong Su https://orcid.org/0000-0003-1104-6015;\\
	Lei Yi https://orcid.org/0000-0002-0247-4600;\\
	Cheng Wang https://orcid.org/0000-0002-6470-7289;\\
	Chao Sun https://orcid.org/0000-0002-0930-6343.}

\appendix

\section{The measurement of droplet size}
\label{sec:append_diameter}
\begin{figure}
    \centerline{\includegraphics[width = 0.8\textwidth]{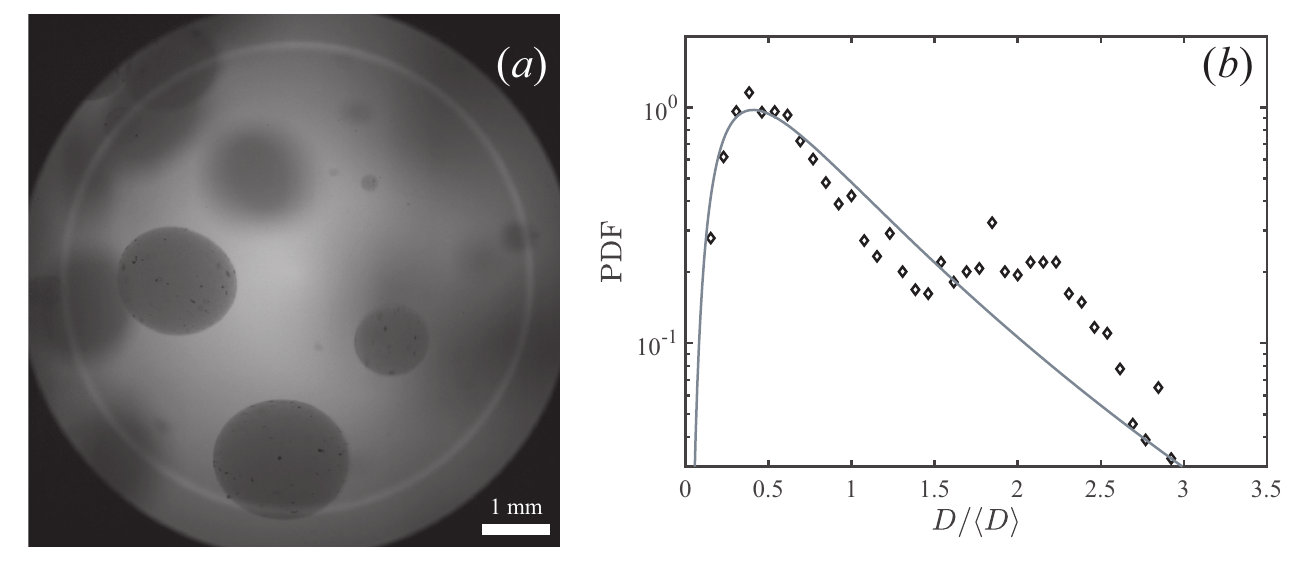}}
    \caption{(\textit{a}) Typical snapshot of the droplets at $\phi = 1\% $ recorded with a high-speed camera connected with a long-distance microscope. (\textit{b}) Probability density function of the normalized droplet size at $\phi=1\%$. The gray solid line represents the log-normal distribution \citep{yi_global_2021}.}
	\label{fig:append_PDF_diam}
\end{figure}

{The droplet size is measured at $\Rey = 1.68 \times 10^4$ with a volume fraction $\phi=1\%$ in the bulk region of TC flow and the average droplet size is $\langle D \rangle \approx 700$ \textmu m. The typical snapshot of droplets and the droplet size distribution are shown in Fig. \ref{fig:append_PDF_diam} (\textit{a}) and (\textit{b}). We also show the log-normal distribution reported by \cite{yi_global_2021}. Our experimental data is consistent with previous study. To obtain the droplet size distribution, one needs a large number of data. At volume fraction of $\phi\geq 5\%$, the interfaces overlap with each other on the captured image, and it is difficult to collect enough data to yield a converged distribution. The detailed droplet size distribution may change at different volume fractions, and this remains an open question to be answered in future study. In this work, we use $\langle D \rangle \approx 700$ \textmu m to represent the mean diameter size in the bulk region at all volume fractions from $\phi = 5\%$ to $\phi = 20\%$.}

{As reported in our previous study \citep{yi_physical_2022}, the mean droplet diameter remains nearly constant when the volume fraction of dispersed phase is less than $50 \%$, which is related to the presence of unavoidable surface-active impurities in the solution suppressing droplet merging. In our previous work \citep{yi_physical_2022} and this study, we use the same flow configuration, and the density and viscosity between the continuous and dispersed phases are nearly matched. The only difference comes from the surface tension: $\gamma \approx 21.1$ and $5.5$ mN/m in this study and previous work, respectively. The value of $\gamma$ will not change the physical mechanism of droplet breakup and coalescence in TC turbulence. Therefore it is reasonable to use $\langle D \rangle \approx 700$ \textmu m to represent the average droplet size at higher volume fractions, which is $0\sim 20\%$ in this study. }
 
{Regarding the radial dependence of droplet size, the TC system can be divided into two parts: the boundary layer near the inner and outer cylinder, the bulk where the flow is nearly homogeneous and isotropic. We would expect smaller droplet size near the boundary layer due to the high shear of the mean velocity \citep{yi_physical_2022}, and a nearly constant droplet size in the bulk. However, as the boundary layer thickness is small, it is difficult to measure the droplet size near the inner cylinder \citep{yi_global_2021}.}

{The radial distribution of droplet number has been investigated by \cite{wang_how_2022,wang_turbulence_2022}. At low Reynolds number, there will be a clustering behaviour near the inner cylinder at the region where the plumes of angular velocity are injected. However, at higher Reynolds number $Re = 1.3 \times 10^4$ which is still lower than the present study, the spatial distribution becomes nearly uniform in the bulk region due to the high turbulence fluctuation.  For our case with higher Reynolds number ($Re =1.48\sim 1.68\times 10^4$), we would expect that the droplets distribution over the radial direction is nearly uniform.}
 
\section{Data processing of velocity time series}
\label{sec:append_data_process}

Regarding the LDA data quality, noise, including stochastic noise from the photodetector and electronics as well as the reflection from the experimental setup, cannot be avoided in the experiment. Following ways were tried to improve the data quality throughout the experiment procedure. Firstly, we set a proper velocity range for each LDA signal channel by properly tuning the frequency shift of the Bragg cell. Secondly, the inner cylinder is anodized to form a black oxidation layer and the upper cover of the setup is made of black Acrylonitrile Butadiene Styrene. The black inner cylinder and upper cover can alleviate laser reflection from solid surface as mentioned in \S \ref{sec:Exp_setup}. Thirdly, for each LDA data series, we used $7$ times the rms value of raw data to filter out the outlier data. The number of outlier data is less than $1$ \textpertenthousand. 

\begin{figure}
    \centering
    \includegraphics[width=1\linewidth]{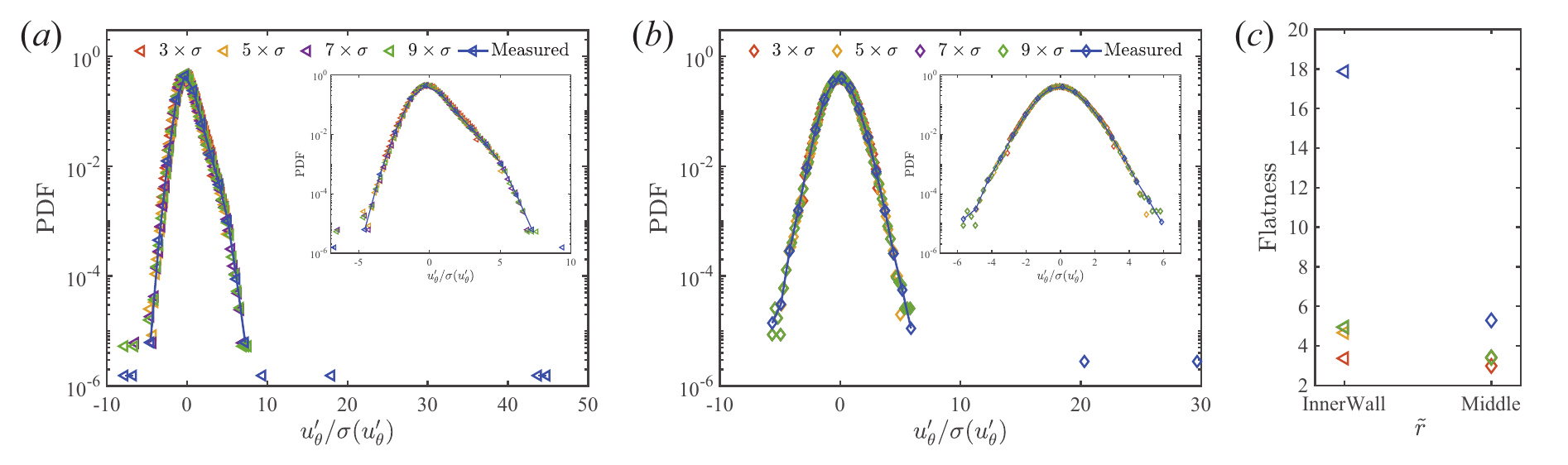}
    \caption{PDFs of times series of $u_\theta$ (\textit{a}) near the inner cylinder and (\textit{b}) at the middle gap using several different thresholds of 3, 5, 7 and 9 times the rms.  (\textit{c}) The corresponding flatness values for each cases. Inset: the zoom-in of the main part of PDFs. }
    \label{fig:append_sigma_number}
\end{figure}
\begin{figure}
	\centerline{\includegraphics[width = 0.6\textwidth]{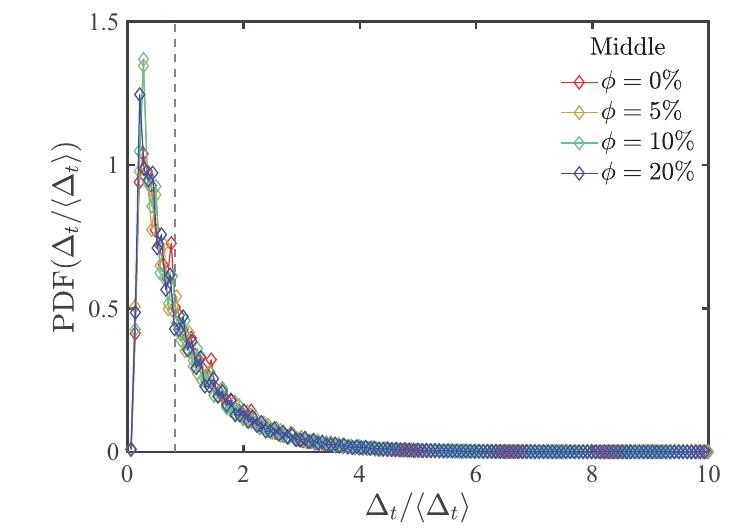}}
	\caption{Probability density functions of the time intervals between two neighbor velocity data points. Here, we use $u_\theta$ measured near the inner cylinder as an example. The vertical dashed line corresponds to the average droplet passing time, i.e. $\langle D \rangle /\langle u_\theta \rangle_{t,z}$.}
	\label{fig:append_delta_t}
\end{figure}

The range of $\pm 7$ times the rms value was selected to eliminate outlier data without severely changing the shape of the probability density function (PDF) as well as the values of flatness. Taking $u_\theta$ at a certain height at $\phi = 20 \% $ as an example, the PDFs and the values of flatness using several different thresholds of 3, 5, 7 and 9 times the rms are shown in Fig. \ref{fig:append_sigma_number}. It reveals outliers (lower right) in the normalized velocity PDFs of $u_\theta$ throughout both the near-wall and middle regions, which are identified as nonphysical artifacts requiring exclusion. The insets in Fig. \ref{fig:append_sigma_number} (\textit{a-b}) demonstrate the filtering performance at different rms multiplier thresholds: while thresholds of 3 and 5 induce artificial truncation of the PDFs tails, values of 7 and 9 exhibit excellent fidelity in preserving the intrinsic flow characteristics while effectively eliminating outlier contamination. Furthermore, flatness analysis reveals a saturation trend with increasing rms multipliers: the flatness values for 7 and 9 become statistically indistinguishable (evidenced by marker overlap in the plot). The discrepancy between raw and processed data flatness originates from the aforementioned invalid outliers. We therefore conclude that the choice of 7 rms value provides the best cutoff without affecting the shape of PDF profiles. 
In our study, the $7\sigma$ threshold yields data exclusion rate smaller than $1$ \textpertenthousand $\ $. While a $1$ \textpertenthousand $\ $ exclusion rate would typically correspond to a $4\sigma$ cutoff in a standard Gaussian distribution, the velocity fluctuations in turbulent flows exhibit deviations from Gaussian statistics due to intermittency. This intermittency leads to elevated tails in PDFs, causing the higher rate of outlier data than a Gaussian distribution.


For LDA measurements, the spatial integration of signals due to the finite size of the measurement volume affects turbulence measurements particularly at the near-wall region \citep{Durst_Jovanović_Sender_1995}. Following their method, the estimations of the magnitude of corrections on $\langle u_\theta \rangle$ and $\sigma(u_\theta ')^2$ are shown below. We denote the relative correction to be 
\begin{equation}
    \Delta_{\langle u_\theta \rangle_{t,z}}/\langle u_\theta \rangle_{t,z} = \frac{d^2}{32}(\frac{\mathrm{d}^2u_{\theta,true}}{\mathrm{d}y^2})/\langle u_\theta \rangle_{t,z},
\end{equation}
\begin{equation}
    \Delta_{\sigma(u^\prime_\theta)^2}/\sigma(u_\theta ')^2 = \frac{d^2}{16}(\frac{\mathrm{d} u_{\theta,true}}{\mathrm{d}y})^2/\sigma(u_\theta ')^2, 
\end{equation}
if we take the first term of Taylor expansion \citep{Durst_Jovanović_Sender_1995}. It can be found that the relative corrections depend only on streamwise velocity gradient at the measurement point.


In the single-phase case, the boundary layer velocity profile obtained by Zhang \textit{et al.} is employed for estimation of the correction. Based on Figure 13 in \citet{Zhang_Fan_Su_Xi_Sun_2025}, the azimuthal velocity profiles in the log layer could be approximately described by $u_\theta^+ = 1/\kappa \ \mathrm{ln} y^+ +B$ at $Re = 1.68 \times 10^5$ and $y^+ = 35 \pm 9$, where $u_\theta^+ =(\omega_ir_i-\langle u_\theta \rangle_{t,z})/u_\tau = (\omega_ir_i-\langle u_\theta \rangle_{t,z})/\sqrt{T/(2\pi \rho_cr_i^2L)},\ y^+ = (r- r_i)/\delta_\nu$ and $\kappa \approx 1$ here (estimated from Figure 13(b) in \cite{Zhang_Fan_Su_Xi_Sun_2025}). Therefore, $\frac{\mathrm{d}^2u_{\theta,true}}{\mathrm{d}y^2} = u_\tau/y^2$ and $(\frac{\mathrm{d} u_{\theta,true}}{\mathrm{d}y})^2 = (u_\tau/y)^2$. Then we have
\begin{equation}
\label{equ:eq_utheta}
    \Delta_{\langle u_\theta \rangle_{t,z}}/\langle u_\theta \rangle_{t,z} = \frac{d^2 u_\tau}{32y^2 \langle u_\theta \rangle_{t,z}}  \approx 6.75 \times 10^{-4},
\end{equation}
\begin{equation}
\label{equ:eq_rms}
    \Delta_{\sigma(u^\prime_\theta)^2}/\sigma(u_\theta ')^2 = \frac{d^2 u_\tau^2}{16y^2 \sigma(u_\theta ')^2} \approx 4.79 \times 10^{-3}. 
\end{equation}
We find that the correction to $\langle u_\theta \rangle_{t,z}$ is smaller than the correction to $\sigma(u_\theta ')^2$, and both corrections are less than $1\%$. This small correction is consistent with previous study \citep{Durst_Jovanović_Sender_1995}. \cite{Durst_Jovanović_Sender_1995} reported that the correction is necessary in the viscous sublayer due to the large velocity gradient, while the correction in the log layer is negligible.

For the two-phase cases at different volume fraction, the mean streamwise velocity profile is not known a priori. \cite{PhysRevFluids.2.083603} have done numerical simulations in wall-bounded turbulent emulsions with volume fraction $\phi = 18.6\%$. They found that the mean streamwise velocity of the two-phase flow shifts slightly upwards with its slope ($1/\kappa$) in the log layer nearly unchanged. Therefore, we can still use Eq. \ref{equ:eq_utheta} and \ref{equ:eq_rms} to calculate the relative corrections. For the two-phase case at the maximum volume fraction ($\phi = 20\%$), $\Delta_{\sigma(u^\prime_\theta)^2}/\sigma(u_\theta ')^2  \approx 7.37 \times 10^{-3}$, which is still less than $1\%$. Consequently, these data acquired is reliable and the effects induced by the finite measurement volume can be considered negligible.

To investigate the effect of droplets on the measured LDA temporal signal, we calculate the PDFs of the time interval, $\Delta_t$, between neighbor velocity data, which is shown in Fig. \ref{fig:append_delta_t}. The horizontal axis is scaled by the average time interval, i.e. $\langle \Delta_t \rangle$. We note that $\langle \Delta_t \rangle$ in the single-phase and two-phase cases are nearly the same and the average data rate is $1/\langle \Delta_t \rangle$. The vertical dashed line denotes the average passing time of one droplet $\langle D \rangle /\langle u_\theta \rangle_{t,z}$. In our experiment, $\langle \Delta_t \rangle> \langle D \rangle /\langle u_\theta \rangle_{t,z}$. It can be found that the PDFs are almost collapsed on the top of each other. Thus, the effect of droplets on the measured temporal signal, if it is present, is only marginal for the current study. One possible explanation for this observation is that   $\langle \Delta_t \rangle > \langle D \rangle /\langle u_\theta \rangle_{t,z}$ and the average spacing between neighbor droplets is larger than $\langle \Delta_t \rangle\langle u_\theta \rangle_{t,z}$ due to the "low" volume fraction. We therefore conclude that the influence of droplets on time series of velocity is negligible.

\bibliographystyle{jfm}

\bibliography{emulsion}

\begin{thebibliography}{76}
\expandafter\ifx\csname natexlab\endcsname\relax\def\natexlab#1{#1}\fi
\def\au#1{#1} \def\ed#1{#1} \def\yr#1{#1}\def\at#1{#1}\def\jt#1{\textit{#1}}
  \def\bt#1{#1}\def\bvol#1{\textbf{#1}} \def\vol#1{#1} \def\pg#1{#1}
  \def\publ#1{#1}\def\arxiv#1{#1}\def\org#1{#1}\def\st#1{\textit{#1}}

\bibitem[Amini \& Hassan(2012)]{amini_investigation_2012}
{\sc \au{Amini, N.} \& \au{Hassan, Y.~A.}} \yr{2012}  \at{An investigation of
  matched index of refraction technique and its application in optical
  measurements of fluid flow}.  \jt{Exp. Fluids}  \bvol{53},  \pg{2011--2020}.

\bibitem[Bakhuis {\em et~al.\/}(2021)Bakhuis, Ezeta, Bullee, Marin, Lohse, Sun
  \& Huisman]{bakhuis_catastrophic_2021}
{\sc \au{Bakhuis, D.}, \au{Ezeta, R.}, \au{Bullee, P.~A.}, \au{Marin, A.},
  \au{Lohse, D.}, \au{Sun, C.} \& \au{Huisman, S.~G.}} \yr{2021}
  \at{Catastrophic phase inversion in high-{R}eynolds-number turbulent
  {T}aylor--{C}ouette flow}.  \jt{Phys. Rev. Lett.}  \bvol{126}~(6),
  \pg{064501}.

\bibitem[Begemann {\em et~al.\/}(2022)Begemann, Trummler, Trautner, Hasslberger
  \& Klein]{begemann2022effect}
{\sc \au{Begemann, A.}, \au{Trummler, T.}, \au{Trautner, E.}, \au{Hasslberger,
  J.} \& \au{Klein, M.}} \yr{2022}  \at{Effect of turbulence intensity and
  surface tension on the emulsification process and its stationary state—{A}
  numerical study}.  \jt{Can. J. Chem. Eng}  \bvol{100}~(12),  \pg{3548--3561}.

\bibitem[Brauckmann \& Eckhardt(2013)]{brauckmann_direct_2013}
{\sc \au{Brauckmann, H.~J.} \& \au{Eckhardt, B.}} \yr{2013}  \at{Direct
  numerical simulations of local and global torque in {T}aylor--{C}ouette flow
  up to {R}e= 30 000}.  \jt{J. Fluid Mech.}  \bvol{718},  \pg{398--427}.

\bibitem[Brauckmann {\em et~al.\/}(2016)Brauckmann, Salewski \&
  Eckhardt]{brauckmann_momentum_2016}
{\sc \au{Brauckmann, H.~J.}, \au{Salewski, M.} \& \au{Eckhardt, B.}} \yr{2016}
  \at{Momentum transport in {T}aylor--{C}ouette flow with vanishing curvature}.
   \jt{J. Fluid Mech.}  \bvol{790},  \pg{419--452}.

\bibitem[Budwig(1994)]{budwig_refractive_1994}
{\sc \au{Budwig, R.}} \yr{1994}  \at{Refractive index matching methods for
  liquid flow investigations}.  \jt{Exp. Fluids}  \bvol{17}~(5),
  \pg{350--355}.

\bibitem[Burin {\em et~al.\/}(2010)Burin, Schartman \& Ji]{burin_local_2010}
{\sc \au{Burin, M.~J.}, \au{Schartman, E.} \& \au{Ji, H.}} \yr{2010}  \at{Local
  measurements of turbulent angular momentum transport in circular {C}ouette
  flow}.  \jt{Exp. Fluids}  \bvol{48},  \pg{763--769}.

\bibitem[Castaing {\em et~al.\/}(1989)Castaing, Gunaratne, Heslot, Kadanoff,
  Libchaber, Thomae, Wu, Zaleski \& Zanetti]{castaing_scaling_1989}
{\sc \au{Castaing, B.}, \au{Gunaratne, G.}, \au{Heslot, F.}, \au{Kadanoff, L.},
  \au{Libchaber, A.}, \au{Thomae, S.}, \au{Wu, X.-Z.}, \au{Zaleski, S.} \&
  \au{Zanetti, G.}} \yr{1989}  \at{Scaling of hard thermal turbulence in
  {R}ayleigh-{B}{\'e}nard convection}.  \jt{J. Fluid Mech.}  \bvol{204},
  \pg{1--30}.

\bibitem[Conan {\em et~al.\/}(2007)Conan, Masbernat, D{\'e}carre \&
  Lin{\'e}]{conan2007local}
{\sc \au{Conan, C.}, \au{Masbernat, O.}, \au{D{\'e}carre, S.} \& \au{Lin{\'e},
  A.}} \yr{2007}  \at{Local hydrodynamics in a dispersed-stratified
  liquid--liquid pipe flow}.  \jt{AICHE J.}  \bvol{53}~(11),  \pg{2754--2768}.

\bibitem[Crialesi-Esposito {\em et~al.\/}(2023{\natexlab{{\em
  a\/}}})Crialesi-Esposito, Boffetta, Brandt, Chibbaro \&
  Musacchio]{crialesi-esposito_intermittency_2023}
{\sc \au{Crialesi-Esposito, M.}, \au{Boffetta, G.}, \au{Brandt, L.},
  \au{Chibbaro, S.} \& \au{Musacchio, S.}} \yr{2023{\natexlab{{\em a\/}}}}
  \at{Intermittency in turbulent emulsions}.  \jt{J. Fluid Mech.}  \bvol{972},
  \pg{A37}.

\bibitem[Crialesi-Esposito {\em et~al.\/}(2024)Crialesi-Esposito, Boffetta,
  Brandt, Chibbaro \& Musacchio]{crialesi-esposito_how_2024}
{\sc \au{Crialesi-Esposito, M.}, \au{Boffetta, G.}, \au{Brandt, L.},
  \au{Chibbaro, S.} \& \au{Musacchio, S.}} \yr{2024}  \at{How small droplets
  form in turbulent multiphase flows}.  \jt{Phys. Rev. Fluids}  \bvol{9}~(7),
  \pg{L072301}.

\bibitem[Crialesi-Esposito {\em et~al.\/}(2023{\natexlab{{\em
  b\/}}})Crialesi-Esposito, Chibbaro \&
  Brandt]{crialesi-esposito_interaction_2023}
{\sc \au{Crialesi-Esposito, M.}, \au{Chibbaro, S.} \& \au{Brandt, L.}}
  \yr{2023{\natexlab{{\em b\/}}}}  \at{The interaction of droplet dynamics and
  turbulence cascade}.  \jt{Commun. Phys.}  \bvol{6}~(1),  \pg{5}.

\bibitem[Crialesi-Esposito {\em et~al.\/}(2022)Crialesi-Esposito, Rosti,
  Chibbaro \& Brandt]{crialesi-esposito_modulation_2022}
{\sc \au{Crialesi-Esposito, M.}, \au{Rosti, M.~E.}, \au{Chibbaro, S.} \&
  \au{Brandt, L.}} \yr{2022}  \at{Modulation of homogeneous and isotropic
  turbulence in emulsions}.  \jt{J. Fluid Mech.}  \bvol{940},  \pg{A19}.

\bibitem[Dodd \& Ferrante(2016)]{dodd_interaction_2016}
{\sc \au{Dodd, M.~S.} \& \au{Ferrante, A.}} \yr{2016}  \at{On the interaction
  of {T}aylor length scale size droplets and isotropic turbulence}.  \jt{J.
  Fluid Mech.}  \bvol{806},  \pg{356--412}.

\bibitem[Dong(2007)]{dong_direct_2007}
{\sc \au{Dong, S.}} \yr{2007}  \at{Direct numerical simulation of turbulent
  {T}aylor--{C}ouette flow}.  \jt{J. Fluid Mech.}  \bvol{587},  \pg{373--393}.

\bibitem[Durst {\em et~al.\/}(1995)Durst, Jovanović \&
  Sender]{Durst_Jovanović_Sender_1995}
{\sc \au{Durst, F.}, \au{Jovanović, J.} \& \au{Sender, J.}} \yr{1995}
  \at{{LDA} measurements in the near-wall region of a turbulent pipe flow}.
  \jt{J. Fluid Mech.}  \bvol{295},  \pg{305–335}.

\bibitem[Eckhardt {\em et~al.\/}(2007{\natexlab{{\em a\/}}})Eckhardt, Grossmann
  \& Lohse]{eckhardt_fluxes_2007}
{\sc \au{Eckhardt, B.}, \au{Grossmann, S.} \& \au{Lohse, D.}}
  \yr{2007{\natexlab{{\em a\/}}}}  \at{Fluxes and energy dissipation in thermal
  convection and shear flows}.  \jt{Europhys. Lett.}  \bvol{78}~(2),
  \pg{24001}.

\bibitem[Eckhardt {\em et~al.\/}(2007{\natexlab{{\em b\/}}})Eckhardt, Grossmann
  \& Lohse]{eckhardt_torque_2007}
{\sc \au{Eckhardt, B.}, \au{Grossmann, S.} \& \au{Lohse, D.}}
  \yr{2007{\natexlab{{\em b\/}}}}  \at{Torque scaling in turbulent
  {T}aylor--{C}ouette flow between independently rotating cylinders}.  \jt{J.
  Fluid Mech.}  \bvol{581},  \pg{221--250}.

\bibitem[Emran \& Schumacher(2008)]{emran_fine-scale_2008}
{\sc \au{Emran, M.~S.} \& \au{Schumacher, J.}} \yr{2008}  \at{Fine-scale
  statistics of temperature and its derivatives in convective turbulence}.
  \jt{J. Fluid Mech.}  \bvol{611},  \pg{13--34}.

\bibitem[Eskin {\em et~al.\/}(2017)Eskin, Taylor \& Yang]{eskin2017modeling}
{\sc \au{Eskin, D.}, \au{Taylor, S.~D.} \& \au{Yang, D.}} \yr{2017}
  \at{Modeling of droplet dispersion in a turbulent {T}aylor--{C}ouette flow}.
  \jt{Chem. Eng. Sci.}  \bvol{161},  \pg{36--47}.

\bibitem[Ezeta {\em et~al.\/}(2018)Ezeta, Huisman, Sun \&
  Lohse]{ezeta2018turbulence}
{\sc \au{Ezeta, R.}, \au{Huisman, S.~G.}, \au{Sun, C.} \& \au{Lohse, D.}}
  \yr{2018}  \at{Turbulence strength in ultimate {T}aylor--{C}ouette
  turbulence}.  \jt{J. Fluid Mech.}  \bvol{836},  \pg{397--412}.

\bibitem[Froitzheim {\em et~al.\/}(2019)Froitzheim, Ezeta, Huisman, Merbold,
  Sun, Lohse \& Egbers]{froitzheim_statistics_2019}
{\sc \au{Froitzheim, A.}, \au{Ezeta, R.}, \au{Huisman, S.~G.}, \au{Merbold,
  S.}, \au{Sun, C.}, \au{Lohse, D.} \& \au{Egbers, C.}} \yr{2019}
  \at{Statistics, plumes and azimuthally travelling waves in ultimate
  {T}aylor--{C}ouette turbulent vortices}.  \jt{J. Fluid Mech.}  \bvol{876},
  \pg{733--765}.

\bibitem[Froitzheim {\em et~al.\/}(2017)Froitzheim, Merbold \&
  Egbers]{froitzheim2017velocity}
{\sc \au{Froitzheim, A.}, \au{Merbold, S.} \& \au{Egbers, C.}} \yr{2017}
  \at{Velocity profiles, flow structures and scalings in a wide-gap turbulent
  {T}aylor--{C}ouette flow}.  \jt{J. Fluid Mech.}  \bvol{831},  \pg{330--357}.

\bibitem[Girifalco \& Good(1957)]{girifalco_theory_1957}
{\sc \au{Girifalco, L.~A} \& \au{Good, R.~J.}} \yr{1957}  \at{A theory for the
  estimation of surface and interfacial energies. {I}. {D}erivation and
  application to interfacial tension}.  \jt{J. Phys. Chem.}  \bvol{61}~(7),
  \pg{904--909}.

\bibitem[Gopalan \& Katz(2010)]{PhysRevLett.104.054501}
{\sc \au{Gopalan, B.} \& \au{Katz, J.}} \yr{2010}  \at{Turbulent shearing of
  crude oil mixed with dispersants generates long microthreads and
  microdroplets}.  \jt{Phys. Rev. Lett.}  \bvol{104},  \pg{054501}.

\bibitem[Grossmann {\em et~al.\/}(2016)Grossmann, Lohse \&
  Sun]{grossmann_highreynolds_2016}
{\sc \au{Grossmann, S.}, \au{Lohse, D.} \& \au{Sun, C.}} \yr{2016}
  \at{High--{R}eynolds number {T}aylor--{C}ouette turbulence}.  \jt{Annu. Rev.
  Fluid Mech.}  \bvol{48}~(1),  \pg{53--80}.

\bibitem[Hinze(1955)]{hinze_fundamentals_1955}
{\sc \au{Hinze, J.~O.}} \yr{1955}  \at{Fundamentals of the hydrodynamic
  mechanism of splitting in dispersion processes}.  \jt{AICHE J.}
  \bvol{1}~(3),  \pg{289--295}.

\bibitem[Hori {\em et~al.\/}(2023)Hori, Ng, Lohse \&
  Verzicco]{hori_interfacial-dominated_2023}
{\sc \au{Hori, N.}, \au{Ng, C.~S.}, \au{Lohse, D.} \& \au{Verzicco, R.}}
  \yr{2023}  \at{Interfacial-dominated torque response in liquid--liquid
  {T}aylor--{C}ouette flows}.  \jt{J. Fluid Mech.}  \bvol{956},  \pg{A15}.

\bibitem[Huisman {\em et~al.\/}(2012{\natexlab{{\em a\/}}})Huisman, van Gils,
  Grossmann, Sun \& Lohse]{huisman_ultimate_2012}
{\sc \au{Huisman, S.~G.}, \au{van Gils, D. P.~M.}, \au{Grossmann, S.}, \au{Sun,
  C.} \& \au{Lohse, D.}} \yr{2012{\natexlab{{\em a\/}}}}  \at{Ultimate
  turbulent {T}aylor--{C}ouette flow}.  \jt{Phys. Rev. Lett.}  \bvol{108}~(2),
  \pg{024501}.

\bibitem[Huisman {\em et~al.\/}(2012{\natexlab{{\em b\/}}})Huisman, van Gils \&
  Sun]{huisman_applying_2012}
{\sc \au{Huisman, S.~G.}, \au{van Gils, D. P.~M.} \& \au{Sun, C.}}
  \yr{2012{\natexlab{{\em b\/}}}}  \at{Applying laser {D}oppler anemometry
  inside a {T}aylor--{C}ouette geometry using a ray-tracer to correct for
  curvature effects}.  \jt{Eur. J. Mech. B/Fluids}  \bvol{36},  \pg{115--119}.

\bibitem[Huisman {\em et~al.\/}(2013{\natexlab{{\em a\/}}})Huisman, Lohse \&
  Sun]{huisman_statistics_2013}
{\sc \au{Huisman, S.~G.}, \au{Lohse, D.} \& \au{Sun, C.}}
  \yr{2013{\natexlab{{\em a\/}}}}  \at{Statistics of turbulent fluctuations in
  counter-rotating {T}aylor--{C}ouette flows}.  \jt{Phys. Rev. E}
  \bvol{88}~(6),  \pg{063001}.

\bibitem[Huisman {\em et~al.\/}(2013{\natexlab{{\em b\/}}})Huisman,
  Scharnowski, Cierpka, K{\"a}hler, Lohse \& Sun]{huisman_logarithmic_2013}
{\sc \au{Huisman, S.~G.}, \au{Scharnowski, S.}, \au{Cierpka, C.},
  \au{K{\"a}hler, C.~J.}, \au{Lohse, D.} \& \au{Sun, C.}}
  \yr{2013{\natexlab{{\em b\/}}}}  \at{Logarithmic boundary layers in strong
  {T}aylor--{C}ouette turbulence}.  \jt{Phys. Rev. Lett.}  \bvol{110}~(26),
  \pg{264501}.

\bibitem[Ibarra {\em et~al.\/}(2021)Ibarra, Matar \&
  Markides]{ibarra_experimental_2021}
{\sc \au{Ibarra, R.}, \au{Matar, O.~K.} \& \au{Markides, C.~N.}} \yr{2021}
  \at{Experimental investigations of upward-inclined stratified oil-water flows
  using simultaneous two-line planar laser-induced fluorescence and particle
  velocimetry}.  \jt{Int. J. Multiph. Flow}  \bvol{135},  \pg{103502}.

\bibitem[Ibarra {\em et~al.\/}(2018)Ibarra, Zadrazil, Matar \&
  Markides]{ibarra_dynamics_2018}
{\sc \au{Ibarra, R.}, \au{Zadrazil, I.}, \au{Matar, O.~K.} \& \au{Markides,
  C.~N.}} \yr{2018}  \at{Dynamics of liquid--liquid flows in horizontal pipes
  using simultaneous two--line planar laser--induced fluorescence and particle
  velocimetry}.  \jt{Int. J. Multiph. Flow}  \bvol{101},  \pg{47--63}.

\bibitem[Kilpatrick(2012)]{kilpatrick2012water}
{\sc \au{Kilpatrick, P.~K.}} \yr{2012}  \at{Water-in-crude oil emulsion
  stabilization: review and unanswered questions}.  \jt{Energy Fuels}
  \bvol{26}~(7),  \pg{4017--4026}.

\bibitem[Kokal(2005)]{kokal2005crude}
{\sc \au{Kokal, S.}} \yr{2005}  \at{Crude-oil emulsions: {A} state-of-the-art
  review}.  \jt{SPE Prod. Fac}  \bvol{20}~(01),  \pg{5--13}.

\bibitem[Kolmogorov(1949)]{kolmogorov_disintegration_1949}
{\sc \au{Kolmogorov, A.~N.}} \yr{1949}  \at{On the disintegration of drops by
  turbulent flows}.  \jt{Dokl. Akad. Nauk SSSR}  \bvol{66},  \pg{825--828}.

\bibitem[Kumara {\em et~al.\/}(2010)Kumara, Halvorsen \&
  Melaaen]{kumara2010particle}
{\sc \au{Kumara, W. A.~S.}, \au{Halvorsen, B.~M.} \& \au{Melaaen, M.~C.}}
  \yr{2010}  \at{Particle image velocimetry for characterizing the flow
  structure of oil--water flow in horizontal and slightly inclined pipes}.
  \jt{Chem. Eng. Sci.}  \bvol{65}~(15),  \pg{4332--4349}.

\bibitem[Kurtz~Jr \& Ward(1936)]{kurtz_refractivity_1936}
{\sc \au{Kurtz~Jr, S.~S.} \& \au{Ward, A.~L.}} \yr{1936}  \at{The refractivity
  intercept and the specific refraction equation of {N}ewton. {I}. development
  of the refractivity intercept and comparison with specific refraction
  equations}.  \jt{J. Frankl. Inst.}  \bvol{222}~(5),  \pg{563--592}.

\bibitem[Lee(1993)]{lee_scope_1993}
{\sc \au{Lee, L.~H.}} \yr{1993}  \at{Scope and limitations of the equation of
  state approach for interfacial tensions}.  \jt{Langmuir}  \bvol{9}~(7),
  \pg{1898--1905}.

\bibitem[Lemenand {\em et~al.\/}(2017)Lemenand, Della~Valle, Dupont \&
  Peerhossaini]{lemenand2017turbulent}
{\sc \au{Lemenand, T.}, \au{Della~Valle, D.}, \au{Dupont, P.} \&
  \au{Peerhossaini, H.}} \yr{2017}  \at{Turbulent spectrum model for
  drop-breakup mechanisms in an inhomogeneous turbulent flow}.  \jt{Chem. Eng.
  Sci.}  \bvol{158},  \pg{41--49}.

\bibitem[Levich(1962)]{levich1962physicochemical}
{\sc \au{Levich, V.~G.}} \yr{1962} {\em Physicochemical Hydrodynamics\/}.
  \publ{Prentice-Hall Inc}.

\bibitem[Li \& Garrett(1998)]{li1998relationship}
{\sc \au{Li, M.} \& \au{Garrett, C.}} \yr{1998}  \at{The relationship between
  oil droplet size and upper ocean turbulence}.  \jt{Mar. Pollut. Bull.}
  \bvol{36}~(12),  \pg{961--970}.

\bibitem[Mandal {\em et~al.\/}(2010)Mandal, Samanta, Bera \&
  Ojha]{mandal2010characterization}
{\sc \au{Mandal, A.}, \au{Samanta, A.}, \au{Bera, A.} \& \au{Ojha, K.}}
  \yr{2010}  \at{Characterization of oil--water emulsion and its use in
  enhanced oil recovery}.  \jt{Ind. Eng. Chem. Res.}  \bvol{49}~(24),
  \pg{12756--12761}.

\bibitem[McClements(2004)]{mcclements2004food}
{\sc \au{McClements, D.~J.}} \yr{2004} {\em Food emulsions: principles,
  practices, and techniques\/}.  \publ{CRC}.

\bibitem[Morgan {\em et~al.\/}(2013)Morgan, Markides, Zadrazil \&
  Hewitt]{morgan2013characteristics}
{\sc \au{Morgan, R.~G.}, \au{Markides, C.~N.}, \au{Zadrazil, I.} \& \au{Hewitt,
  G.~F.}} \yr{2013}  \at{Characteristics of horizontal liquid--liquid flows in
  a circular pipe using simultaneous high-speed laser-induced fluorescence and
  particle velocimetry}.  \jt{Int. J. Multiph. Flow}  \bvol{49},  \pg{99--118}.

\bibitem[Mukherjee {\em et~al.\/}(2019)Mukherjee, Safdari, Shardt,
  Kenjere{\v{s}} \& Van~den Akker]{mukherjee_dropletturbulence_2019}
{\sc \au{Mukherjee, S.}, \au{Safdari, A.}, \au{Shardt, O.}, \au{Kenjere{\v{s}},
  S.} \& \au{Van~den Akker, H. E.~A.}} \yr{2019}  \at{Droplet--turbulence
  interactions and quasi-equilibrium dynamics in turbulent emulsions}.  \jt{J.
  Fluid Mech.}  \bvol{878},  \pg{221--276}.

\bibitem[Newton(1704)]{newton_opticks_1704}
{\sc \au{Newton, I.}} \yr{1704} {\em Opticks: Or, {A} treatise of the
  reflections, refractions, inflexions and colours of light\/}.  \publ{S.
  Smith, and B. Walford}.

\bibitem[Ni(2024)]{ni2024deformation}
{\sc \au{Ni, R.}} \yr{2024}  \at{Deformation and breakup of bubbles and drops
  in turbulence}.  \jt{Annu. Rev. Fluid Mech.}  \bvol{56}~(1),  \pg{319--347}.

\bibitem[Perlekar(2019)]{perlekar2019kinetic}
{\sc \au{Perlekar, P.}} \yr{2019}  \at{Kinetic energy spectra and flux in
  turbulent phase-separating symmetric binary-fluid mixtures}.  \jt{J. Fluid
  Mech.}  \bvol{873},  \pg{459--474}.

\bibitem[Perlekar {\em et~al.\/}(2014)Perlekar, Benzi, Clercx, Nelson \&
  Toschi]{perlekar_spinodal_2014}
{\sc \au{Perlekar, P.}, \au{Benzi, R.}, \au{Clercx, H. J.~H.}, \au{Nelson,
  D.~R.} \& \au{Toschi, F.}} \yr{2014}  \at{Spinodal decomposition in
  homogeneous and isotropic turbulence}.  \jt{Phys. Rev. Lett.}
  \bvol{112}~(1),  \pg{014502}.

\bibitem[Perlekar {\em et~al.\/}(2012)Perlekar, Biferale, Sbragaglia,
  Srivastava \& Toschi]{perlekar_droplet_2012}
{\sc \au{Perlekar, P.}, \au{Biferale, L.}, \au{Sbragaglia, M.}, \au{Srivastava,
  S.} \& \au{Toschi, F.}} \yr{2012}  \at{Droplet size distribution in
  homogeneous isotropic turbulence}.  \jt{Phys. Fluids}  \bvol{24}~(6).

\bibitem[Piela {\em et~al.\/}(2008)Piela, Delfos, Ooms, Westerweel \&
  Oliemans]{piela2008phase}
{\sc \au{Piela, K.}, \au{Delfos, R.}, \au{Ooms, G.}, \au{Westerweel, J.} \&
  \au{Oliemans, R. V.~A.}} \yr{2008}  \at{On the phase inversion process in an
  oil--water pipe flow}.  \jt{Int. J. Multiph. Flow}  \bvol{34}~(7),
  \pg{665--677}.

\bibitem[Pope(2000)]{Pope_2000}
{\sc \au{Pope, S.~B.}} \yr{2000} {\em Turbulent Flows\/}.  \publ{Cambridge
  University Press}.

\bibitem[Procaccia {\em et~al.\/}(1991)Procaccia, Ching, Constantin, Kadanoff,
  Libchaber \& Wu]{procaccia_transitions_1991}
{\sc \au{Procaccia, I.}, \au{Ching, E. S.~C}, \au{Constantin, P.},
  \au{Kadanoff, L.~P.}, \au{Libchaber, A.} \& \au{Wu, X.-Z.}} \yr{1991}
  \at{Transitions in convective turbulence: the role of thermal plumes}.
  \jt{Phys. Rev. A}  \bvol{44}~(12),  \pg{8091}.

\bibitem[Reis {\em et~al.\/}(2010)Reis, Lampreia, Santos, Moita \&
  Douh{\'e}ret]{reis_refractive_2010}
{\sc \au{Reis, J. C.~R.}, \au{Lampreia, I. M.~S.}, \au{Santos, {\^A}. F.~S.},
  \au{Moita, M. L. C.~J.} \& \au{Douh{\'e}ret, G.}} \yr{2010}  \at{Refractive
  index of liquid mixtures: theory and experiment}.  \jt{ChemPhysChem}
  \bvol{11}~(17),  \pg{3722--3733}.

\bibitem[Risso \& Fabre(1998)]{risso1998oscillations}
{\sc \au{Risso, F.} \& \au{Fabre, J.}} \yr{1998}  \at{Oscillations and breakup
  of a bubble immersed in a turbulent field}.  \jt{J. Fluid Mech.}  \bvol{372},
   \pg{323--355}.

\bibitem[Roccon {\em et~al.\/}(2017)Roccon, De~Paoli, Zonta \&
  Soldati]{PhysRevFluids.2.083603}
{\sc \au{Roccon, A.}, \au{De~Paoli, M.}, \au{Zonta, F.} \& \au{Soldati, A.}}
  \yr{2017}  \at{Viscosity-modulated breakup and coalescence of large drops in
  bounded turbulence}.  \jt{Phys. Rev. Fluids}  \bvol{2},  \pg{083603}.

\bibitem[Rosti {\em et~al.\/}(2019)Rosti, Ge, Jain, Dodd \&
  Brandt]{rosti_droplets_2019}
{\sc \au{Rosti, M.~E.}, \au{Ge, Z.}, \au{Jain, S.~S.}, \au{Dodd, M.~S.} \&
  \au{Brandt, L.}} \yr{2019}  \at{Droplets in homogeneous shear turbulence}.
  \jt{J. Fluid Mech.}  \bvol{876},  \pg{962--984}.

\bibitem[Shang {\em et~al.\/}(2003)Shang, Qiu, Tong \& Xia]{shang2003measured}
{\sc \au{Shang, X.-D.}, \au{Qiu, X.-L.}, \au{Tong, P.} \& \au{Xia, K.-Q.}}
  \yr{2003}  \at{Measured local heat transport in turbulent rayleigh-b{\'e}nard
  convection}.  \jt{Phys. Rev. Lett.}  \bvol{90}~(7),  \pg{074501}.

\bibitem[Spernath \& Aserin(2006)]{spernath2006microemulsions}
{\sc \au{Spernath, A.} \& \au{Aserin, A.}} \yr{2006}  \at{Microemulsions as
  carriers for drugs and nutraceuticals}.  \jt{Adv. Colloid Interface Sci.}
  \bvol{128},  \pg{47--64}.

\bibitem[Su {\em et~al.\/}(2024{\natexlab{{\em a\/}}})Su, Wang, Zhang, Xu, Wang
  \& Sun]{su_turbulence_2024}
{\sc \au{Su, J.}, \au{Wang, C.}, \au{Zhang, Y.-B.}, \au{Xu, F.}, \au{Wang, J.}
  \& \au{Sun, C.}} \yr{2024{\natexlab{{\em a\/}}}}  \at{Turbulence modulation
  in liquid–liquid two-phase {T}aylor–{C}ouette turbulence}.  \jt{J. Fluiid
  Mech.}  \bvol{999},  \pg{A98}.

\bibitem[Su {\em et~al.\/}(2024{\natexlab{{\em b\/}}})Su, Yi, Zhao, Wang, Xu,
  Wang \& Sun]{su_numerical_2024}
{\sc \au{Su, J.}, \au{Yi, L.}, \au{Zhao, B.}, \au{Wang, C.}, \au{Xu, F.},
  \au{Wang, J.} \& \au{Sun, C.}} \yr{2024{\natexlab{{\em b\/}}}}  \at{Numerical
  study on the mechanism of drag modulation by dispersed drops in two-phase
  {T}aylor--{C}ouette turbulence}.  \jt{J. Fluiid Mech.}  \bvol{984},  \pg{R3}.

\bibitem[Su {\em et~al.\/}(2025)Su, Zhang, Wang, Yi, Xu, Fan, Wang \&
  Sun]{Su_Zhang_Wang_Yi_Xu_Fan_Wang_Sun_2025}
{\sc \au{Su, J.}, \au{Zhang, Y.-B.}, \au{Wang, C.}, \au{Yi, L.}, \au{Xu, F.},
  \au{Fan, Y.}, \au{Wang, J.} \& \au{Sun, C.}} \yr{2025}  \at{How interfacial
  tension enhances drag in turbulent {T}aylor–{C}ouette flow with neutrally
  buoyant and equally viscous droplets}.  \jt{J. Fluid Mech.}  \bvol{1002},
  \pg{A2}.

\bibitem[Trefftz-Posada \& Ferrante(2023)]{trefftz-posada_interaction_2023}
{\sc \au{Trefftz-Posada, P.} \& \au{Ferrante, A.}} \yr{2023}  \at{On the
  interaction of {T}aylor length-scale size droplets and homogeneous shear
  turbulence}.  \jt{J. Fluid Mech.}  \bvol{972},  \pg{A9}.

\bibitem[Wang {\em et~al.\/}(2023)Wang, DeGroot \&
  Floryan]{wang_numerical_2023}
{\sc \au{Wang, C.}, \au{DeGroot, C.~T.} \& \au{Floryan, J.~M.}} \yr{2023}
  \at{Numerical simulation of drag reduction for turbulent flow in cylindrical
  annuli with axial corrugations}.  \jt{Trans. Can. Soc. Mech. Eng.}
  \bvol{48}~(1),  \pg{164--172}.

\bibitem[Wang {\em et~al.\/}(2022{\natexlab{{\em a\/}}})Wang, Yi, Jiang \&
  Sun]{wang_how_2022}
{\sc \au{Wang, C.}, \au{Yi, L.}, \au{Jiang, L.} \& \au{Sun, C.}}
  \yr{2022{\natexlab{{\em a\/}}}}  \at{How do the finite-size particles modify
  the drag in {T}aylor--{C}ouette turbulent flow}.  \jt{J. Fluid Mech.}
  \bvol{937},  \pg{A15}.

\bibitem[Wang {\em et~al.\/}(2022{\natexlab{{\em b\/}}})Wang, Yi, Jiang \&
  Sun]{wang_turbulence_2022}
{\sc \au{Wang, C.}, \au{Yi, L.}, \au{Jiang, L.} \& \au{Sun, C.}}
  \yr{2022{\natexlab{{\em b\/}}}}  \at{Turbulence drag modulation by dispersed
  droplets in {T}aylor--{C}ouette flow: the effects of the dispersed phase
  viscosity}.  \jt{J. Fluid Mech.}  \bvol{952},  \pg{A39}.

\bibitem[Wiederseiner {\em et~al.\/}(2011)Wiederseiner, Andreini, Epely-Chauvin
  \& Ancey]{wiederseiner_refractive-index_2011}
{\sc \au{Wiederseiner, S.}, \au{Andreini, N.}, \au{Epely-Chauvin, G.} \&
  \au{Ancey, C.}} \yr{2011}  \at{Refractive-index and density matching in
  concentrated particle suspensions: a review}.  \jt{Exp. Fluids}  \bvol{50},
  \pg{1183--1206}.

\bibitem[Wright {\em et~al.\/}(2017)Wright, Zadrazil \&
  Markides]{wright_review_2017}
{\sc \au{Wright, S.~F.}, \au{Zadrazil, I.} \& \au{Markides, C.~N.}} \yr{2017}
  \at{A review of solid--fluid selection options for optical-based measurements
  in single-phase liquid, two-phase liquid--liquid and multiphase solid--liquid
  flows}.  \jt{Exp. Fluids}  \bvol{58},  \pg{1--39}.

\bibitem[Yakhot(1989)]{yakhot_probability_1989}
{\sc \au{Yakhot, V.}} \yr{1989}  \at{Probability distributions in
  high-{R}ayleigh number {B}{\'e}nard convection}.  \jt{Phys. Rev. Lett.}
  \bvol{63}~(18),  \pg{1965}.

\bibitem[Yi {\em et~al.\/}(2024)Yi, Girotto, Toschi \& Sun]{yi2024divergence}
{\sc \au{Yi, L.}, \au{Girotto, I.}, \au{Toschi, F.} \& \au{Sun, C.}} \yr{2024}
  \at{Divergence of critical fluctuations on approaching catastrophic phase
  inversion in turbulent emulsions}.  \jt{Phys. Rev. Lett.}  \bvol{133}~(13),
  \pg{134001}.

\bibitem[Yi {\em et~al.\/}(2021)Yi, Toschi \& Sun]{yi_global_2021}
{\sc \au{Yi, L.}, \au{Toschi, F.} \& \au{Sun, C.}} \yr{2021}  \at{Global and
  local statistics in turbulent emulsions}.  \jt{J. Fluid Mech.}  \bvol{912},
  \pg{A13}.

\bibitem[Yi {\em et~al.\/}(2023)Yi, Wang, Huisman \& Sun]{yi2023recent}
{\sc \au{Yi, L.}, \au{Wang, C.}, \au{Huisman, S.~G.} \& \au{Sun, C.}} \yr{2023}
   \at{Recent developments of turbulent emulsions in {T}aylor--{C}ouette flow}.
   \jt{Philos. Trans. R. Soc. A}  \bvol{381}~(2243),  \pg{20220129}.

\bibitem[Yi {\em et~al.\/}(2022)Yi, Wang, van Vuren, Lohse, Risso, Toschi \&
  Sun]{yi_physical_2022}
{\sc \au{Yi, L.}, \au{Wang, C.}, \au{van Vuren, T.}, \au{Lohse, D.}, \au{Risso,
  F.}, \au{Toschi, F.} \& \au{Sun, C.}} \yr{2022}  \at{Physical mechanisms for
  droplet size and effective viscosity asymmetries in turbulent emulsions}.
  \jt{J. Fluid Mech.}  \bvol{951},  \pg{A39}.

\bibitem[Zhang {\em et~al.\/}(2025)Zhang, Fan, Su, Xi \&
  Sun]{Zhang_Fan_Su_Xi_Sun_2025}
{\sc \au{Zhang, Y.-B.}, \au{Fan, Y.}, \au{Su, J.}, \au{Xi, H.-D.} \& \au{Sun,
  C.}} \yr{2025}  \at{Global drag reduction and local flow statistics in
  {T}aylor–{C}ouette turbulence with dilute polymer additives}.  \jt{J. Fluid
  Mech.}  \bvol{1002},  \pg{A33}.

\end{thebibliography}

\end{document}